%% file: main.tex
\documentclass[manuscript,table]{acmart} 
\usepackage{packages}
\renewcommand\footnotetextcopyrightpermission[1]{}

\AtBeginDocument{%
  \providecommand\BibTeX{{%
    \normalfont B\kern-0.5em{\scshape i\kern-0.25em b}\kern-0.8em\TeX}}}

\begin{document}

\title{A Methodological Framework for LLM-Based Mining of Software Repositories}

\begin{abstract}
Large Language Models (LLMs) are increasingly used in software engineering research, offering new opportunities for automating repository mining tasks. However, despite their growing popularity, the methodological integration of LLMs into Mining Software Repositories (MSR) remains poorly understood. Existing studies tend to focus on specific capabilities or performance benchmarks, providing limited insight into how researchers utilize LLMs across the full research pipeline. To address this gap, we conduct a mixed-method study that combines a rapid review and questionnaire survey in the field of LLM4MSR. We investigate (1) the approaches and (2) the threats that affect the empirical rigor of researchers involved in this field. Our findings reveal 15 methodological approaches, nine main threats, and 25 mitigation strategies. Building on these findings, we present PRIMES 2.0, a refined empirical framework organized into six stages, comprising 23 methodological substeps, each mapped to specific threats and corresponding mitigation strategies, providing prescriptive and adaptive support throughout the lifecycle of LLM-based MSR studies. Our work contributes to establishing a more transparent and reproducible foundation for LLM-based MSR research.
\end{abstract}
\author{Vincenzo De Martino}
\email{vdemartino@unisa.it}
\orcid{0000-0003-1485-4560}
\affiliation{%
  \institution{University of Salerno}
  \city{Salerno}
  \country{Italy}
}

\author{Joel Castaño}
\email{joel.castano@upc.edu}
\orcid{0000-0002-XXXX-XXXX}
\affiliation{%
  \institution{Universitat Politècnica de Catalunya}
  \city{Barcelona}
  \country{Spain}
}

\author{Fabio Palomba}
\email{fpalomba@unisa.it}
\orcid{0000-0001-9337-5116}
\affiliation{%
  \institution{University of Salerno}
  \city{Salerno}
  \country{Italy}
}

\author{Xavier Franch}
\email{xavier.franch@upc.edu}
\orcid{0000-0002-XXXX-XXXX}
\affiliation{%
  \institution{Universitat Politècnica de Catalunya}
  \city{Barcelona}
  \country{Spain}
}

\author{Silverio Martínez-Fernández}
\email{silverio.martinez@upc.edu}
\orcid{0000-0003-XXXX-XXXX}
\affiliation{%
  \institution{Universitat Politècnica de Catalunya}
  \city{Barcelona}
  \country{Spain}
}

\renewcommand{\shortauthors}{De Martino et al.}

\date{Received: date / Accepted: date}

\maketitle

\input{Section/Introduction}
\input{Section/Background}

\input{Section/Methodology}
\input{Section/Results}

\input{Section/Discussion}
\input{Section/Threats}
\input{Section/Conclusion}

\section*{Data Availability Statement}
This paper includes supplementary data provided in the online appendix. The datasets generated from the rapid review and survey, along with the raw data, detailed plots, and additional resources necessary to reproduce our analysis, are available at: \cite{appendix}.

\section*{Credits}
\textbf{Vincenzo De Martino}: Conceptualization, Methodology, Formal analysis, Investigation, Data Curation, Validation, Project administration, Writing - Original Draft, Writing - Review \& Editing, Visualization.
\textbf{Joel Castaño}: Conceptualization, Validation, Formal analysis, Investigation, Writing - Original Draft, Writing - Review \& Editing, Visualization.
\textbf{Fabio Palomba}:
Conceptualization, Methodology, Writing - Review \& Editing, Supervision.
\textbf{Xavier Franch}: Conceptualization, Methodology, Writing - Review \& Editing, Supervision.
\textbf{Silverio Martínez-Fernández}: Conceptualization, Methodology, Writing - Review \& Editing, Supervision.

\section*{Acknowledgments}
During the preparation of this work, the author utilized ChatGPT to enhance the language and readability. After using this service, the authors thoroughly reviewed and edited the content as necessary and take full responsibility for the publication's content. This work has been partially supported by the \textsl{GAISSA-Optimizer} research project under the AGAUR 2025 program (Code: PROD 00236), and by the \textsl{EVOLUCIÓN CONTINUA Y EFICIENTE DE SISTEMAS DE ML: UN ENFOQUE DIRIGIDO POR EL ECOSISTEMA} Spanish research project (Code: PID2024-156019OB-I00).

\bibliographystyle{ACM-Reference-Format}  
\bibliography{bib/bib}

\end{document}

%% file: Section/Introduction.tex
\section{Introduction}
\label{sec:introduction}
Large Language Models (LLMs) have taken on a central role in Software Engineering (SE), reshaping established practices in software development, maintenance, and comprehension \cite{hou2024large,gonzalez2025pre, fan2023large}. Recent work by Hou et al. \cite{hou2024large} has systematically mapped how LLMs are used in SE, providing a comprehensive taxonomy of tasks supported by LLMs across the development lifecycle, from code generation \cite{xu2024licoeval,dong2024self} and comprehension \cite{nam2024using} to testing \cite{pourasad2024does}, and maintenance \cite{pomian2024next}. Their findings highlight the breadth of LLM applicability and their disruptive potential in automating and augmenting main SE practices.

While recent work has mapped where LLMs can be applied across Software Engineering tasks, less attention has been given to how LLMs are used as part of the software engineering research process itself. \textbf{Rather than dealing with the use of LLMs to assist software engineers, this article focuses on understanding and improving the methodological use of LLMs by SE researchers conducting empirical studies.} In particular, we focus on Mining Software Repositories (MSR) as a domain where LLMs are increasingly employed \cite{hou2024large}. Recent contributions have begun to explore this methodological shift, examining how LLMs are used to support systematic reviews~\cite{felizardo2025difficulties,costalonga2025can} and qualitative analysis~\cite{lecca2025applications,barros2025large}. Building on these efforts, our work provides an overview of how researchers operationalize LLMs in MSR and identifies approaches and threats in this new emerging research area.

Traditionally, MSR has relied on hand-crafted rules \cite{de4960056into,tufano2015and}, manual analysis \cite{hindle2008large,chen2020comprehensive,castano2023exploring}, and classical machine learning methods \cite{ugurel2002s,mcmillan2011categorizing,bangash2019developers} to extract insights from development artifacts such as commits, issues, and pull requests. These approaches typically focused on structured data and required extensive manual annotation and data preparation \cite{calefato2022preliminary,bernardo2024machine}. Today, the advent of LLMs is reshaping how repository mining is approached. Thanks to their ability to process unstructured and heterogeneous sources of information, LLMs can synthesize knowledge across multiple artifacts. Unlike traditional MSR pipelines, which relied on task-specific models and explicit feature engineering, LLM-based workflows are increasingly driven by prompt design, model configuration, and iterative validation without the need to conduct extensive pre-processing steps \cite{gonzalez2025pre,de2024developers,sheikhaei2024empirical}. In the survey of Hou et al. \cite{hou2024large}, the MSR topic is briefly acknowledged as a use case for LLMs; however, the study, like much of the existing literature, remains task-centric and overlooks how LLMs are operationalized in practice. Remarkably, using LLMs for MSR introduces a new set of threats. Researchers must design prompts tailored to heterogeneous artifacts, manage non-deterministic and context-sensitive model outputs, and validate results across large datasets without standardized protocols. Yet, these workflows are rarely described in sufficient detail. Decisions regarding data sampling, model selection, inference configuration, and evaluation strategies are often implicit, making replication difficult and methodological rigor inconsistent.

Recent studies have begun to build a methodological foundation for using LLMs in research. These efforts range from proposing high-level guidelines for evaluation and responsible use \cite{sallou2024breaking, wagner2024towards}, to empirically testing LLM capabilities for specific research activities like study replication and results validation \cite{liang2024can, wang2025can}, and introducing preliminary frameworks for MSR tasks \cite{de2025framework}. 

\steattentionboxa{\faExclamationTriangle \hspace{0.05cm}
While recent studies have introduced high-level guidelines and preliminary frameworks for using LLMs in empirical software engineering, they often remain abstract, focusing on evaluation principles or individual tasks. What is still missing is a systematic, bottom-up understanding of how researchers make concrete design decisions throughout the entire LLM-based MSR pipeline. This lack of grounded operational insight makes it difficult to assess whether current approaches effectively minimize threats to validity, and limits our ability to build reproducible and methodologically sound studies.} 

This study aims to fill a gap by developing a methodological framework that captures how LLMs are used in MSR workflows. The framework is grounded in an empirical investigation that identifies methodological approaches, threats to validity, and the design choices employed to ensure rigor and reproducibility. Building on our previous work~\cite{de2025framework}, which introduced a preliminary version of the PRIMES framework based on our experience, this study extends and generalizes those insights through a broader empirical investigation. Complementing prior efforts to propose LLM-oriented guidelines~\cite{wagner2024towards, sallou2024breaking} and reflect on the methodological implications of LLM usage in SE research~\cite{felizardo2025difficulties, trinkenreich2025get, barros2025large}, our study offers a domain-specific contribution focused on LLM-based repository mining.

To better understand how researchers employ LLM in MSR, we conducted a mixed-method exploratory study that combines a rapid review of 15 (out of initially considered 7,973) peer-reviewed articles \cite{cartaxo2020rapid} with a survey with 22 (initially targeted over 143) researchers experienced in LLM-based mining studies. This investigation revealed that the integration of LLMs into MSR is reshaping the set of activities that researchers conduct, changing methodological approaches and introducing new threats to validity that must be carefully addressed throughout the research process. Yet, in the absence of a structured and empirical framework, many studies risk falling short of the rigor required for reproducibility and validity. When researchers apply LLMs without clear objectives, well-designed prompts, or systematic validation, results may appear plausible but reflect shallow or inconsistent reasoning. Without methodological rigor, studies can unintentionally produce misleading findings, draw unsupported conclusions, or generate outputs that cannot be replicated. Until such frameworks become standard, there remains a high risk that LLM-based MSR research continues to grow in volume but not in reliability. To address this gap, we refine and extend our initial framework~\cite{de2025framework} by introducing PRIMES 2.0, a structured methodological guide designed to support the responsible and rigorous use of LLMs in MSR studies. PRIMES 2.0 defines a sequence of six actionable stages across the research workflow, while explicitly highlighting the potential threats that may arise if these substages are neglected or poorly implemented. For each threat, the framework proposes concrete mitigation strategies, offering researchers practical guidance to ensure reproducibility, interpretability, and methodological soundness. Beyond a static checklist, we envision PRIMES 2.0 as an evolving resource that can guide current researchers and be iteratively improved as tools, models, and empirical needs evolve over time.

\smallskip
\noindent \textbf{Structure of the Paper.} 
Section~\ref{sec:related} presents the background and related work on MSR and the emerging use of LLMs in this domain. It examines how researchers are currently operationalizing LLMs in practice, particularly within MSR, and identifies the research gap that motivates our investigation.
Section \ref{sec:methodology} introduces our research questions and details the methodology adopted in our mixed-method exploratory study.
Section \ref{sec:results} reports the findings derived from both the rapid review and the questionnaire survey. Section \ref{sec:threats} discusses the threats to validity associated with our study design and data analysis.
Section \ref{sec:discussion} reflects on the implications of our results, providing a synthesis of emerging practices and open threats consolidated within our PRIMES 2.0 framework.
Finally, Section \ref{sec:conclusion} concludes the paper by outlining avenues for future research.

%% file: Section/Background.tex
\section{Background and Related Work}
\label{sec:related}
In this section, we present the background and empirical context that motivate our study. First, we introduce the PRIMES framework, which is a preliminary artifact designed to guide the use of Large Language Models (LLMs) in repository mining studies (MSR). The PRIMES framework provides a foundation for understanding the various phases, practices, and risks associated with empirical research on LLM-based software engineering (SE).

Next, we describe existing literature on the use of LLMs in empirical software engineering, focusing on recent studies that have proposed evaluation guidelines, identified threats to validity, and explored the capabilities of these models in research contexts.

\subsection{PRIMES: Prompt Refinement and Insights for Mining Empirical Software repositories}
\label{sec:primes}
To structure the methodological use of LLMs in repository mining research, we introduced in a previous work the PRIMES framework—Prompt Refinement and Insights for Mining Empirical Software Repositories~\cite{de2025framework}. PRIMES was developed as an initial response to the growing need for reproducible and empirical practices when using LLMs for automated data collection in Mining Software Repositories (MSR). The framework offers a structured approach for designing, executing, and validating LLM-based MSR studies, combining empirical observations with methodological guidance. PRIMES is composed of four main stages. 

\textbf{`Stage 1: Creation of Prompts for Piloting'} focuses on defining clear research objectives, selecting prompting strategies, such as one-shot, few-shot, chain-of-thought, or structured prompting, and specifying the desired output format to facilitate analysis and interpretation. \textbf{`Stage 2: Prompt Pilot Test'} introduces a systematic process for evaluating prompt quality through dual annotation, where outputs generated by the LLM are compared to human annotations using statistical measures such as Cohen’s kappa. Prompts are iteratively refined until the agreement reaches a predefined threshold. \textbf{`Stage 3: Evaluation Among Multiple LLMs'} addresses the selection of the most appropriate model for a given MSR task. It involves constructing a reliable oracle, benchmarking different LLMs based on metrics such as accuracy, interpretability, and cost, and assessing the generalizability of prompts across models. Finally, \textbf{`stage 4: Output Validation'} ensures the reliability and traceability of LLM-generated outputs by detecting hallucinations, enforcing consistent formatting, tracking provenance, and automating validation.

Each stage reflects lessons learned from prior empirical studies and is designed to address threats such as prompt sensitivity, validation complexity, and model variability. Originally conceived as a preliminary checklist to guide LLM-based MSR research, PRIMES provides a practical foundation for ensuring methodological rigor and reproducibility.
However, PRIMES was based only on two of our own previous LLM-based MSR studies; therefore, the empirical baseline of the framework remained preliminary. In response, this study builds on PRIMES to develop PRIMES 2.0, an extended version informed by a mixed-method investigation of current practices. PRIMES 2.0 incorporates empirically grounded refinements derived from both literature and practitioner experience, aiming to support more robust and transparent empirical SE research.

\subsection{Related Work}
The advent of LLMs has profoundly impacted the SE landscape. A growing body of research, synthesized in recent surveys~\cite{hou2024large, fan2023large}, demonstrates the widespread application of LLMs across the software development lifecycle. These studies showcase LLMs' capabilities for tasks such as automated code generation~\cite{xu2024licoeval, dong2024self}, code comprehension and summarization~\cite{nam2024using}, test generation and fault localization~\cite{pourasad2024does}, and software maintenance and refactoring~\cite{pomian2024next}. The primary focus of these contributions is typically on assessing LLMs performance in executing specific SE tasks. As such, most of the literature to date approaches LLMs from an application perspective, evaluating their effectiveness, accuracy, or efficiency in concrete use cases. While this line of work has been instrumental in demonstrating what LLMs can do, fewer studies have addressed how LLMs should be methodologically integrated into research pipelines, particularly in the context of empirical software engineering and MSR. This distinction between application and methodological focus motivates our investigation.

To address these concerns, recent work has started outlining empirical guidelines for researchers using LLMs in SE. Wagner et al.~\cite{wagner2024towards}, for instance, propose one of the most comprehensive sets of methodological guidelines to date. Their guidelines categorizes LLM-related research into three types: (i) studies where LLMs are tools for empirical data collection or annotation, (ii) studies where LLMs are the subject of evaluation, and (iii) observational studies analyzing developer interactions with LLMs. For each category, the authors propose practical recommendations aimed at ensuring transparency, reproducibility, and interpretability. These include clearly stating the LLM provider, model version, temperature and decoding settings, and the date of access; documenting the structure and rationale of the prompt used; and justifying the chosen output evaluation metrics. Wagner et al.~\cite{wagner2024towards} also emphasize the need for human validation, reporting inter-annotator agreement, and distinguishing between LLM errors and human misjudgments.

Sallou et al.~\cite{sallou2024breaking} complement this perspective by identifying threats to validity when conducting LLM-based empirical studies. They draw attention to three major threats: the volatility and non-replicability of closed-source LLMs, the risk of unintentional data leakage from the model’s training corpus, and the lack of observability into the internal decision-making process of LLMs. These factors compromise both internal and external validity, making it difficult to compare results over time or across research teams. As mitigation, the authors recommend designing experiments with prompt versioning, including metadata for traceability, and using obfuscation techniques on input data to limit the potential influence of memorized training content. While these guidelines are not empirically validated in specific domains such as MSR, they provide valuable direction for structuring future research protocols. Our study complements this work by identifying existing practices in the field and organizing them into distinct stages.

Moving from theoretical guidelines to threat to validity, Felizardo et al.~\cite{felizardo2025difficulties} offer an in-depth account of the difficulties encountered when using LLMs to assist in conducting and replicating Systematic Literature Reviews (SLRs). In two case studies, they evaluate ChatGPT’s support for automating the screening of titles and abstracts. Their findings reveal a number of concrete issues that hinder the adoption of LLMs in SLR workflows. First, they observe high prompt sensitivity, where minor rewordings of the same instruction produce significantly different outputs. Second, they document how randomness in model responses, combined with unclear memory of prior context, undermines reproducibility. Third, they find that enforcing structured inclusion/exclusion criteria via prompt engineering is particularly difficult, especially when LLMs are fed limited textual input. The authors argue that while LLMs can augment certain SLR activities, their deployment must be accompanied by rigorous methodological controls and community standards for prompt reporting and output validation.

Related concerns emerge in the context of qualitative empirical research. Leça et al.~\cite{lecca2025applications} explore the use of LLMs for supporting qualitative data analysis, such as thematic coding, categorization, and clustering. Based on a case study with empirical SE data, they highlight both the benefits and limitations of LLMs in interpretive tasks. On the one hand, LLMs offer time-saving advantages, especially for less experienced researchers, and can surface plausible labels or groupings. On the other hand, their outputs often lack the nuance, contextual sensitivity, and reflexivity required in qualitative research. More problematically, LLMs may invent codes or themes that reflect generic or statistically plausible associations rather than grounded interpretations of the data. The authors caution that LLMs should not be treated as automated replacements for domain experts, and call for methodological frameworks that embed LLMs into qualitative workflows in a structured, auditable, and human-in-the-loop manner.

Similar concerns are echoed by Kausar et al.~\cite{ahmed2025can}, who investigate whether LLMs can effectively replace manual annotation in software engineering tasks such as commit labeling and issue classification. Their findings reveal that, although LLMs can sometimes approximate human annotation quality, they frequently produce inconsistent or superficial results, especially when prompts are underspecified or the input context is ambiguous. The study highlights risks, including semantic drift, prompt sensitivity, and inadequate transparency in output rationale. The authors conclude that without robust validation strategies and clearly defined prompt designs, LLMs cannot reliably serve as substitutes for manual annotation. 

In parallel to the development of empirical guidelines and threat models, a growing number of position papers and empirical explorations have reflected more critically on the broader epistemological and methodological implications of integrating LLMs into SE research. Trinkenreich et al.~\cite{trinkenreich2025get} provide a high-level framework for interpreting the transformative role of LLMs across the entire research pipeline, applying McLuhan’s Tetrad of Media Effects (Enhance, Obsolesce, Retrieve, Reverse). They argue that LLMs are not merely tools for productivity but media that fundamentally reshape how research is conceived, conducted, and communicated. Their analysis emphasizes that LLMs can enhance creativity (e.g., by accelerating ideation and hypothesis generation), but may also lead to undesirable reversals, such as homogenization of research questions and creativity echo chambers, if adopted uncritically. Importantly, they advocate for maintaining human agency and intellectual ownership in all LLM-supported research workflows, calling for the development of community-wide standards, shared benchmarks, and educational resources.

Complementary insights come from Barros et al.~\cite{barros2025large}, who focus on the use of LLMs in qualitative research within SE. Their systematic mapping reveals that LLMs are increasingly used to support tasks such as open coding, thematic analysis, and categorization of interview transcripts or issue tracker comments. While these applications show promise in reducing manual effort and accelerating the analysis process, the authors also identify critical limitations. Notably, LLMs often lack the domain sensitivity and contextual awareness needed to capture the subtleties of developer reasoning or project-specific nuances. Furthermore, their output may reflect superficial associations, fail to capture researcher reflexivity, or introduce subtle biases due to training data artifacts. Barros et al.~\cite{barros2025large}, recommend human-in-the-loop workflows, transparent reporting of prompt strategies, and post-hoc validation techniques to ensure that qualitative insights remain grounded and trustworthy.

These studies illustrate LLMs potential to augment empirical SE research at scale, but their methodological integration raises complex threats that cannot be fully addressed by high-level principles alone. They reinforce the importance of designing concrete, domain-specific frameworks that account for the nuances of how LLMs are used in practice.

However, despite the richness of the emerging discourse, a notable gap persists. Most existing studies either focus on a single type of task, provide top-down recommendations without empirical grounding, or address SE research at a general level without tailoring their findings to the specifics of MSR. For instance, while the threats of reproducibility and prompt sensitivity are acknowledged, there is limited guidance on how to operationalize solutions within the MSR domain.

\novelty{While prior work focuses on LLM capabilities or general methodological principles, our study examines how LLMs are operationalized in MSR research. By synthesizing insights from the literature and researchers, we identify recurring methodological approaches and threats to validity, along with strategies to mitigate them. The targeted result is PRIMES 2.0, a structured and empirically grounded framework that supports rigorous, reproducible, and threat-aware LLM-based repository mining.}

%% file: Section/Methodology.tex
\begin{figure}[h]
    \centering
    \includegraphics[width=1\linewidth]{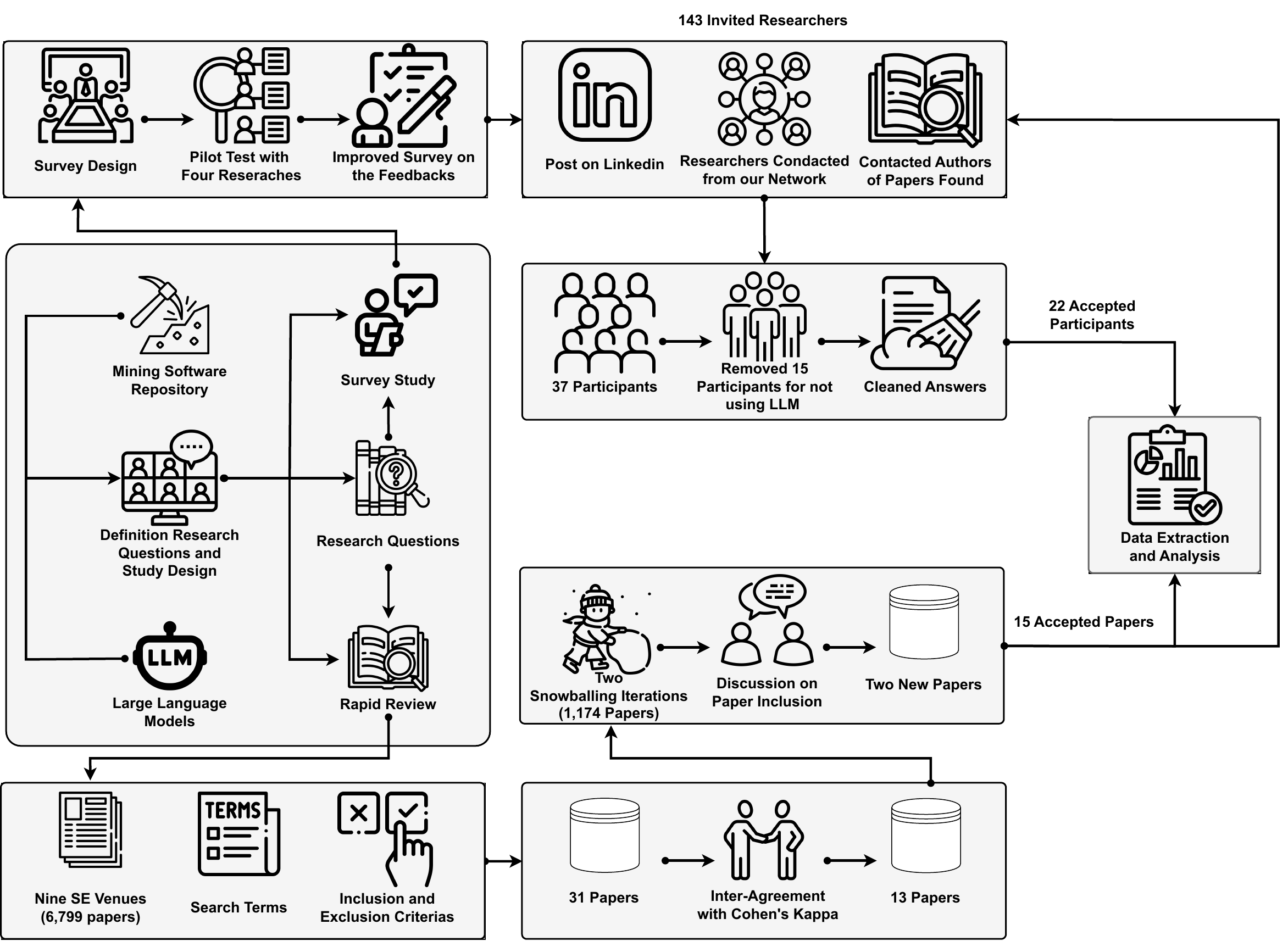}
    \caption{Overview of the Research Process.}
    \label{fig:overview}
\end{figure}

\section{Research Process Overview}
\label{sec:methodology}
To define our research goal, we follow the Goal-Question-Metric (GQM) approach \cite{caldiera1994goal}. 
The \emph{goal} of this study is to \emph{analyze} the use of large language models (LLMs) \emph{for the purpose of} data collection and analysis \emph{with respect to} empirical rigor and threats to validity mitigation \emph{from the perspective of} of SE researchers \emph{in the context of} mining software repository (MSR).
Researchers aim to understand recurring obstacles and usage patterns to inform future tools and research agendas and seek actionable insights to integrate LLMs into MSR workflows effectively.

To address this objective, we focus on two main research questions (\textbf{RQ}s). First, \textbf{RQ$_1$} aims to understand how LLMs are used in practice for MSR. This includes the specific patterns employed, prompting techniques, validation strategies, and configuration choices that determine how LLMs function in research scenarios \cite{de2025framework,hou2024large}. Understanding these usage patterns enables the identification of common practices, tool configurations, and validation strategies, which can support the development of practical guidelines, improve reproducibility, and foster more effective integration of LLMs into MSR workflows. Therefore, we ask:

\sterqbox{RQ\textsubscript{1}—Large Language Model Approaches for Empirical Rigor}{What approaches are employed when using large language models for repository mining?}

Subsequently, \textbf{RQ$_2$} investigates the specific threats to validity encountered when applying LLMs to MSR. While LLMs are increasingly employed in various tasks, they also introduce new technical, methodological, and cost-related difficulties \cite{hou2024large,sallou2024breaking}. By identifying these threats, we aim to highlight the issues that can inform future mitigation strategies and empirical validation efforts. Therefore, we ask:

\sterqbox{RQ\textsubscript{2}—Large Language Model Threats}{What are the current threats to validity faced by researchers when they use large language models for repository mining?}

To answer these research questions, we adopted a \textit{mix-method} exploratory study \cite{johnson2007toward} consisting of two complementary empirical studies. Specifically, we conducted a rapid review to analyze existing published papers and a survey study to collect insights from researchers. Although the two studies differ in scope and methodology, they both contribute to answering the same research questions by capturing evidence from both the published literature and contemporary practitioner experience.

The first empirical study consisted of a \textit{rapid review} of peer-reviewed literature \cite{cartaxo2020rapid}, designed to systematically collect and synthesize existing studies describing how LLMs are applied in repository mining studies. We followed the guidelines of Kitchenham et al. \cite{kitchenham2009systematic}  and scoped our search to the period between 2020 and 2025, as informed by Hou et al. \cite{hou2024large}. The search targeted high-impact software engineering venues (e.g., ICSE, MSR, TOSEM), and applied inclusion and exclusion criteria focused on empirical applications of LLMs to repository mining studies. To increase coverage and identify additional relevant studies potentially missed during the initial query, we also conducted backward and forward snowballing on the references and citations of the selected papers \cite{10.1145/2601248.2601268}. This process led to the identification of 15 relevant studies. While informative, the limited number of retrieved articles, despite the growing relevance of LLMs in SE, indicated that the academic literature alone might not offer a sufficiently broad or up-to-date picture of current practices in the field.

To address this limitation, we complemented the review with a second empirical study: a structured \textit{questionnaire survey study} aimed at collecting insights from researchers experienced in LLM-based mining of software repositories. The survey followed best practices for empirical research, as recommended by Kitchenham et al.~\cite{kitchenham2008_PersonalOpinionSurveys} and Dillman et al.~\cite{dillman2014internet}. The survey included both closed and open-ended questions to capture respondents’ prompting strategies, validation practices, tool usage, configuration choices, and perceived threats. We conducted a pilot test with four researchers to refine clarity and validity, then distributed the final survey among academic and industry researchers working on MSR with LLMs. This allowed us to gather a more diverse set of quantitative and qualitative responses that extend the findings of the rapid review with the practitioner perspectives. The integration of these two methods improves our ability to triangulate evidence effectively. The review provides a comprehensive overview of established practices, drawing on an extensive published literature for a solid foundation. The survey investigates the current landscape, capturing current and novel insights from practitioners. This dual approach enriches current understanding and provides a more nuanced perspective on the use of LLMs in repository mining studies.

To ensure methodological rigor and reporting transparency, we followed the ACM/SIGSOFT Empirical Standards\footnote{\url{https://github.com/acmsigsoft/EmpiricalStandards}}, particularly the guidelines under the \textsl{``Review Articles''}, \textsl{``Questionnaire Surveys''}, and \textsl{``Mixed Methods''} categories. In the remainder of this section, we describe each empirical study in detail, as illustrated in Figure~\ref{fig:overview}.

\subsection{Rapid Review}
\label{sec:review}
We conducted a \textit{rapid review}, a methodology that streamlines or omits certain steps of a traditional SLR to synthesize evidence in a shortened timeframe \cite{cartaxo2020rapid}. The rationale for choosing this approach over a standard SLR is twofold. First, the application of LLMs in software engineering is a nascent and rapidly evolving field; a systematic literature review (SLR) would be time-consuming and risk becoming obsolete before publication, failing to capture the most current practices. Second, our goal is to provide a timely summary of this emerging landscape to identify current threats and inform our subsequent investigation. As a result, we focused on SE venues that publish articles on repository mining studies and articles that have been published up to 2025. A rapid review is expressly designed for such contexts. While reducing the number of possible articles that we may have omitted, they offer a compromise that prioritizes timeliness and relevance to a contemporary topic rather than exhaustiveness. Throughout the process, we still followed established guidelines by Kitchenham et al. \cite{kitchenham2009systematic} where applicable to maintain methodological rigor.


\subsubsection{Seed Set Search}
We constructed our search strategy following the established guidelines by Kitchenham et al. \cite{kitchenham2009systematic} for identifying primary studies in empirical SE. To define our search space, we adopted a \textit{seed set approach}, a commonly used strategy in studies targeting emerging technologies (e.g., \cite{hou2024large, de2025classification}). Our goal was to identify peer-reviewed research articles that (i) explicitly utilize LLMs in the SE context, and (ii) apply these models specifically to tasks related to MSR.

The search process was carried out in three structured steps:

\smallskip
\textbf{Publication Period.} In line with prior literature on LLM usage in SE \cite{hou2024large}, we scoped our review to the period between 2020 and 2025. This interval reflects the emergence and maturation of generative models (e.g., GPT-3, Claude 3 Haiku), and aligns with the recent growth of empirical studies leveraging LLMs in practice.

\smallskip
\textbf{Target Venues.} We selected nine high-quality SE venues known for publishing empirical research, as shown in Table~\ref{table:list_venue}, analyzing a total of 6,799 papers in the selected venues.

\begin{table}[ht]
    \caption{Publication Venues.}
    \label{table:list_venue}

    \rowcolors{2}{gray!15}{white}
    \resizebox{\linewidth}{!}{
    \begin{tabular}{|c|p{15cm}|} 
        \hline 
        \rowcolor{black}
        \textcolor{white}{\textbf{Venue}} & \textcolor{white}{\textbf{Name}} \\
        \hline
        Journal & ACM Transactions on Software Engineering and Methodology (TOSEM)\\
        Journal & IEEE Transactions on Software Engineering (TSE)\\
        Journal & Empirical Software Engineering (EMSE)\\
        Conference & International Conference on Software Engineering (ICSE)\\
        Conference & Mining Software Repositories (MSR)\\
        Conference & International Conference on Automated Software Engineering (ASE) \\
        Conference & Joint European Software Engineering Conference and Symposium on the Foundations of Software Engineering (ESEC/FSE) \\
        Conference & ACM SIGSOFT International Symposium on Software Testing and Analysis (ISSTA) \\
        Conference & Empirical Software Engineering and Measurement (ESEM)  \\
        \hline
    \end{tabular}
    }
\end{table}

These venues were chosen based on their relevance to empirical studies and their increasing inclusion of LLM-related research in recent years. By concentrating on these sources, we ensured coverage of high-quality, peer-reviewed contributions relevant to our research questions.

\smallskip
\textbf{Search Terms and Fields.} We applied our search terms to the \textit{title}, \textit{abstract}, \textit{conclusion} and \textit{keywords} fields of each article \cite{kitchenham2009systematic}. The query was constructed using Boolean logic: we used the \texttt{OR} operator to include alternative expressions and synonymous terms within each thematic group, and the \texttt{AND} operator to connect conceptually distinct groups—ensuring that retrieved studies addressed both the use of LLMs and their application to software repository mining.

The terms were derived from our research questions and informed by recent terminology used in the literature by Hou et al. \cite{hou2024large}, who identified common descriptors in LLM-related software engineering studies. To address variations in terminology to address our goal on the repository mining studies, we narrowed our search to include acronyms, synonyms, and both general and specific terms. We grouped the terms into two main categories:

\begin{itemize}
    \item \textbf{LLM-related keywords:} ``Large Language Model'', ``LLM'', ``Generative AI'', and names of specific LLMs such as ``GPT'', ``Claude'', ``Gemini'', ``LLaMA''.

    \item \textbf{Software repository mining keywords:} ``Empirical Software Engineering'', ``Automated Data Collection'', ``Mining Software Repository'', ``Repository mining'', ``Open-Source Project'', as well as names of specific open-source repositories such as ``GitHub'', ``Hugging Face'', and ``StackOverflow''.
\end{itemize}

This query structure was designed to maximize both \textit{recall}, by covering lexical variation, and \textit{precision}, by filtering for conceptual relevance. The resulting set of papers was subsequently screened using the predefined inclusion and exclusion criteria to identify the final set of studies analyzed in this review.

\subsubsection{Exclusion and Inclusion Criteria}
\begin{table}[h]
    \caption{Inclusion and exclusion criteria used in the selection process.}
    \label{tab:inclusion_exclusion}
    \rowcolors{1}{gray!15}{white}
    \resizebox{\linewidth}{!}{
    \begin{tabular}{p{15cm}}
        \hline
        \rowcolor{black}
        \textcolor{white}{\textbf{Inclusion Criteria (IC)}} \\
        \hline
        (IC1) The article explicitly states that at least one Large Language Model (LLM) is used. \\
        (IC2) The study applies LLMs to software repository-related tasks (e.g., commit classification, data extraction). \\
        (IC3) The study presents empirical results, practical experience, or methodological contributions. \\
        \hline
        \rowcolor{black}
        \textcolor{white}{\textbf{Exclusion Criteria (EC)}} \\
        \hline
        (EC1) Duplicate articles or multiple versions of similar studies authored by the same group. \\
        (EC2) Studies published in books, theses, monographs, keynotes, panels, or venues without a peer-review process. \\
        (EC3) Tool demonstrations, editorials, or grey literature (e.g., technical reports, preprints). \\
        (EC4) Articles not written in English. \\
        (EC5) Articles that mention LLMs but do not describe their application to software repository analysis. \\
        (EC6) Studies focused on LLM applications unrelated to repository mining (e.g., code generation, summarization, literature review). \\
        \hline
    \end{tabular}
    }
\end{table}
To ensure the methodological rigor and relevance of our review, we defined a set of inclusion and exclusion criteria prior to the screening process. These criteria were informed by established practices in empirical software engineering~\cite{kitchenham2009systematic} and aligned with the approach adopted by recent studies on LLMs in SE \cite{hou2024large,zheng2025towards}. Given our objective to analyze how LLMs are used for MSR, the criteria were designed to filter in only those studies that offer direct, documented evidence of such applications, while systematically excluding irrelevant or low-quality sources.

We included studies that (i) explicitly employ at least one Large Language Model (LLM), and (ii) apply that model to tasks directly related to software repositories, including mining activities, commit classification, quality analysis, or automated data extraction. We also required that selected studies provide empirical results or methodological guidance within the scope of LLM-based repository mining. In contrast, we excluded studies that lacked technical depth, methodological clarity, or contextual relevance. In particular, we excluded duplicate or overlapping publications from the same research, studies published in non-peer-reviewed articles (e.g., theses, preprints, book chapters, editorials), and papers not written in English\footnote{These two last ECs apply when considering snowballing, see next subsection}. Furthermore, we excluded studies that merely referenced LLMs without providing a description of their application to repository mining, as well as studies focusing on software engineering techniques to improve LLMs, rather than on the application of LLMs to software engineering problems.

Unlike Hou et al. \cite{hou2024large}, we limited our scope to peer-reviewed articles to ensure consistency and scientific validity of the selected literature. The full list of inclusion and exclusion criteria applied during the review is summarized in Table \ref{tab:inclusion_exclusion}.

The screening process was conducted by the first author, who applied the defined inclusion and exclusion criteria to the initial set of 6,799 papers. As a result, 31 primary studies were selected.

\subsubsection{Study Selection Validation.} Following the initial screening process, the first and second authors independently reviewed the 31 selected papers to assess their eligibility for final inclusion. To quantify their agreement, we employed Cohen’s Kappa coefficient \cite{cohen1960coefficient}, using a three-point agreement scale: 0 indicating disagreement, 0.5 indicating the need for further discussion, and 1 indicating full agreement. A Cohen’s Kappa value of 0.75 or higher was considered acceptable, indicating substantial agreement. In the first round, the resulting Kappa value was 0.57, indicating moderate agreement. To resolve discrepancies and ensure consistent application of the inclusion criteria, the authors conducted a consensus meeting to discuss all cases of disagreement and clarify ambiguous interpretations. A second round of evaluation was then performed, which resulted in a perfect agreement score (Cohen’s Kappa = 1.00).

As a result of this process, 18 papers were excluded, leaving a final set of 13 primary studies. To enhance the coverage of relevant literature, we subsequently conducted a backward and forward snowballing phase on the reference lists and citations of the included papers.

\subsubsection{Snowballing.} To complement the seed set and ensure broader coverage of relevant literature, the first author conducted a backward and forward snowballing process, following the guidelines of Wohlin et al. \cite{10.1145/2601248.2601268}. The process was conducted using \textsl{Google Scholar},\footnote{Link to \textsl{Google Scholar}: \url{https://scholar.google.com/}} an academic search engine that facilitates the analysis by allowing rapid access to both referenced and citing works.

The snowballing was executed in two levels. In the first level, the references and citations of the 13 included studies were reviewed, resulting in a total of 991 papers screened (876 Backward and 115 Forward). Two papers were identified as potentially relevant in this phase. The second level involved applying the same process to the outputs of the first-level snowballing, during which an additional 183 papers were reviewed (174 Backward and 9 Forward); however, no further papers met the inclusion criteria. The two papers identified were reviewed in a joint meeting by the first and second authors. After discussion, both papers were judged to meet the inclusion criteria and were approved for inclusion in the final dataset. These additions brought the total number of included primary studies to 15.

\subsubsection{Rapid Review Data Extraction and Analysis.} 
\begin{figure}[htbp]
  \centering
  \begin{minipage}[b]{0.45\textwidth}
    \includegraphics[width=\textwidth]{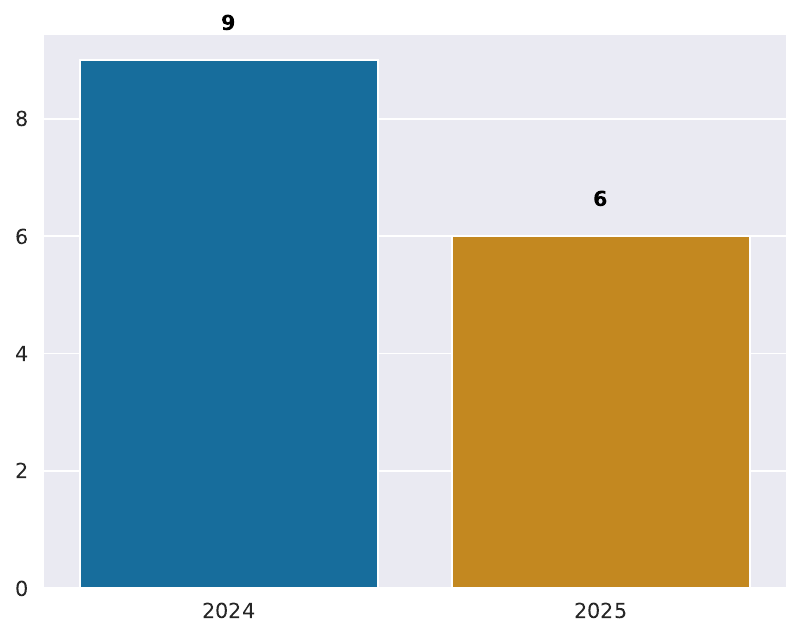}
    \caption{Papers published over time.}
    \label{fig:year_paper}
  \end{minipage}
  \begin{minipage}[b]{0.45\textwidth}
    \includegraphics[width=\textwidth]{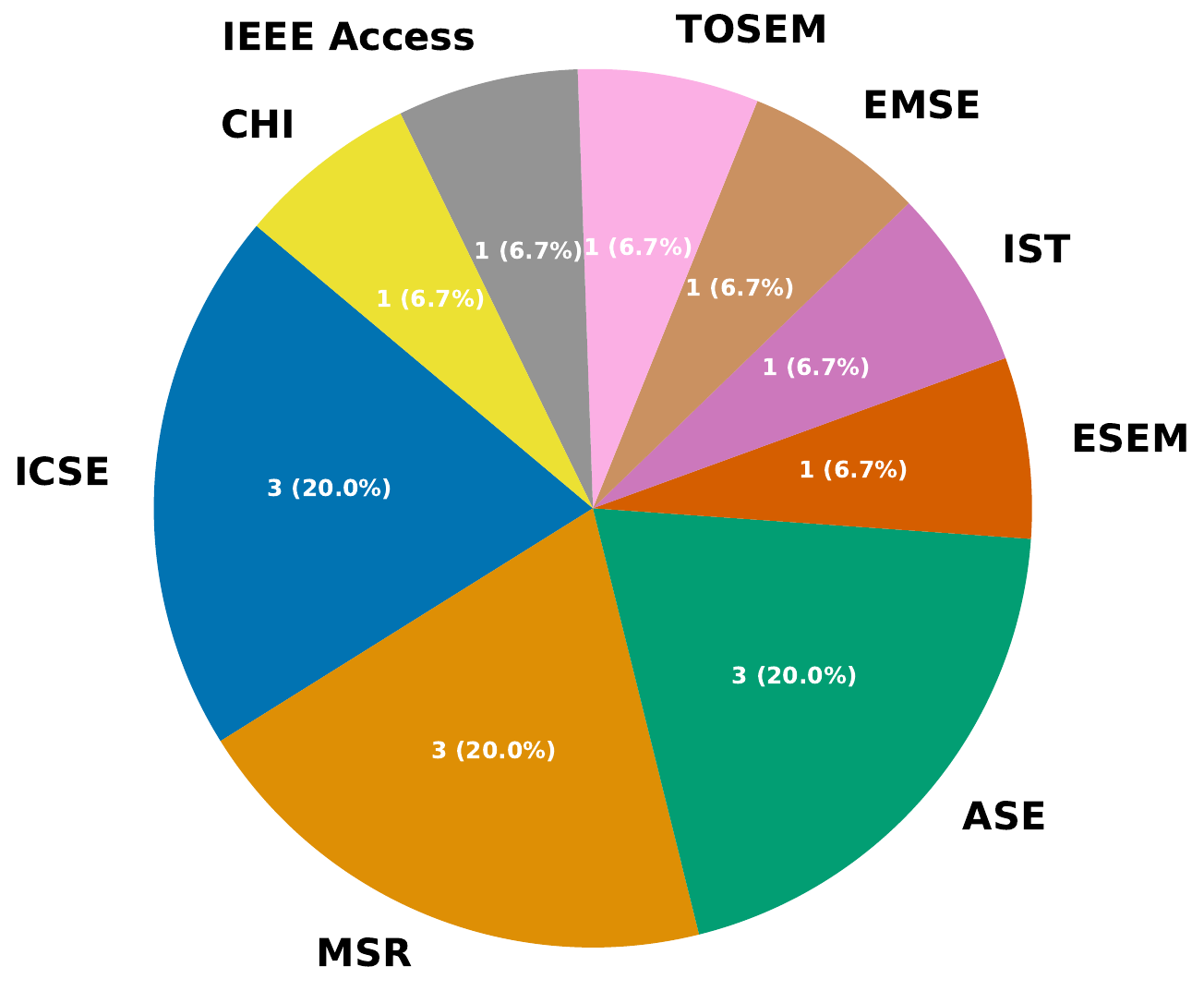}
    \caption{Papers venues.}
    \label{fig:venue_paper}
  \end{minipage}
\end{figure}
To support the synthesis of results and ensure consistent evidence collection across studies, we designed a structured data extraction protocol. The first author performed the data extraction using a predefined spreadsheet, and the second author verified the consistency and completeness of the extracted fields. Each extracted attribute was defined prior to analysis and aligned with our research questions.

We first extracted general metadata to support the descriptive and bibliographic analysis of the selected corpus. As shown in Figure~\ref{fig:year_paper}, the distribution by year of the selected studies reflects the recent emergence of LLM usage in software engineering: no papers were published before 2024, nine studies in 2024 and six published in 2025. This confirms that the integration of LLMs into repository mining is a very recent phenomenon.

We also analyzed the types of venues in which the studies appeared. As illustrated in Figure~\ref{fig:venue_paper}, the corpus includes 10 conference papers and five journal articles. This balance reflects both the exploratory nature of the topic, often first reported in conference venues, and its growing academic maturity, as evidenced by the presence of journal publications.

For \textbf{RQ\textsubscript{1}}, which investigates the practices and approaches adopted for applying LLMs to MSR, we collected methodological and technical details about the application of LLMs. Specifically, we recorded the primary focus of the study (e.g., issue classification, Self-Admitted technical debt detection), the software repositories analyzed (e.g., GitHub, Hugging Face), and the LLM employed, distinguishing between proprietary (e.g., GPT, Claude) and open-source LLMs (e.g., LLaMA, Mistral). We also extracted detailed information about the prompt engineering strategies used, including the prompting pattern (e.g., zero-shot, few-shot, chain-of-thought), techniques, and structural elements (e.g., objective, task). Additionally, we captured the activities performed using LLMs, whether the authors followed or referred to existing LLM usage guidelines, and how the output of the model was validated, such as through manual annotation, ground truth comparison, or statistical agreement. Evaluation metrics (e.g., Accuracy, F1-score, API cost) and configuration parameters (e.g., temperature, max tokens, top-$k$) were also recorded to characterize reproducibility.

To address \textbf{RQ\textsubscript{2}}, which focuses on threats and limitations in using LLMs for MSR, we extracted all information explicitly describing encountered barriers or risks. This included threats related to model behavior (e.g., hallucinations, instability, or high cost), methodological concerns (e.g., reproducibility or scalability), and any threats to validity reported by the articles. We paid particular attention to issues related to the trustworthiness and generalizability of LLM-generated outputs, which are often cited concerns in LLM4SE studies.

\subsection{Questionnaire Survey}
Simultaneously, we conducted a \textit{questionnaire survey study} to complement the evidence gathered through the rapid review and capture insights from researchers actively working with LLMs in software repository mining. Following established guidelines for empirical surveys in software engineering~\cite{kitchenham2008_PersonalOpinionSurveys, dillman2014internet}, we designed our instrument to ensure clarity, reliability, and methodological rigor. The questionnaire survey approach was particularly suited to our objective of exploring how LLMs are used in practice, given the fast-evolving nature of the field and the limitations of relying solely on published literature to capture emerging threats and practices.
\subsubsection{Design of the Online Questionnaire Survey}

In designing the questionnaire survey study, we adhered to the guidelines outlined by Dillman et al. \cite{dillman2014internet} and Kitchenham et al. \cite{kitchenham2008_PersonalOpinionSurveys}. Dillman et al.\cite{dillman2014internet} tailored design method offers a structured framework for improving both response rates and the overall quality of questionnaire survey data, emphasizing principles such as clear question formulation, personalized communication, and respondent engagement. In parallel, Kitchenham et al.\cite{kitchenham2008_PersonalOpinionSurveys} empirical guidance for conducting surveys in software engineering informed our decisions regarding questionnaire structure, sampling strategies, and mitigation of bias.

To uphold ethical standards and ensure participant anonymity, we provided an informed consent statement at the beginning of the survey. The questionnaire survey design was reviewed and approved by the Ethical Committee of the University of Salerno. We deliberately avoided collecting any personally identifiable information such as name, email, gender, or age. Instead, we focused on collecting role-specific and organizational context data to characterize respondents' experience levels. To further increase the reliability of our dataset, the questionnaire survey instrument included embedded attention check questions to identify and later exclude responses provided carelessly or without engagement.

To validate the questionnaire survey, we employed an iterative pilot-testing process aimed at ensuring clarity, completeness, and accurate estimation of response time. The pilot was conducted with four participants, each having at least one year of experience working with LLM technologies in a software engineering context. These individuals, distinct from those participating in the final study, were selected for their domain relevance and ability to offer constructive critique. Their feedback was used to refine the questionnaire by eliminating redundant items and correcting typographical issues.\footnote{More specifically, some redundant questions were removed, and typos that were identified were corrected.} We also measured the average time required to complete the survey, which was about 10 minutes, which is well below our goal of balancing coverage and burden on respondents.

\subsubsection{Structure of the Questionnaire Survey}
\label{sec:survey_structure}

The questionnaire survey was designed as a single instrument comprising five thematic sections, with two sections mapped to our research objectives. The structure balanced closed-ended questions, enabling quantitative analysis, with open-ended questions to capture deeper insights into participants’ practices and threats. The full list of survey questions is provided in the online appendix \cite{appendix}.

\begin{enumerate}
    \item \textbf{Consent and Background Information:} The questionnaire survey began with an informed consent form outlining the study’s purpose, confidentiality measures, and voluntary participation conditions. Participants were required to confirm their consent before proceeding. The section then gathered information on the respondent’s current role in research or industry, experience with empirical software engineering, and previous involvement in LLM-based repository mining. Respondents also specified which LLMs and software repositories they had worked with (e.g., GPT-4, Claude, GitHub, StackOverflow). An attention check was included to filter careless responses. To ensure that only relevant respondents proceeded, we included a screening question: “Have you conducted studies on the use of LLMs for software repository analysis?” Participants who answered “No” were automatically redirected to the end of the questionnaire survey, and their responses were not recorded.

    \smallskip
    \item \textbf{Practices in LLM-based Repository Mining:} This section, aligned with RQ\textsubscript{2}, focused on how participants apply LLMs in practice. It covered prompting techniques (e.g., zero-shot, few-shot, chain-of-thought), types of tasks performed, prompt structure, evaluation metrics, automation of pipeline components, and whether established guidelines or best practices were followed. Participants were also asked to describe how they validate LLM outputs and how they configure model parameters (e.g., temperature, top-$k$).

    \smallskip
    \item \textbf{Threats and Limitations:} This section addressed RQ\textsubscript{1} and focused on practical and methodological obstacles encountered when using LLMs. Participants rated the frequency of various threats, such as hallucinations, high cost, reproducibility issues, and scalability limitations, and were asked to identify the most significant threats they had faced. Open-ended questions invited them to elaborate on mitigation strategies and threats to validity they had observed.

    \smallskip
    \item \textbf{Perceptions and Future Needs:} This section asked participants to rate, using a five-point Likert scale, the potential usefulness of structured guidelines for improving LLM-based repository mining. Respondents were also asked to suggest tools, resources, or forms of support they believed would be most beneficial in improving their current practices.

    \smallskip
    \item \textbf{Final Reflections:} The questionnaire survey concluded with an open-text question inviting participants to share any additional experiences, insights, or observed bad practices related to the use of LLMs in software repository contexts. The final screen displayed a short note of appreciation for their time and contribution.
\end{enumerate}

\subsubsection{Questionnaire Survey Sample and Procedure}

For participant selection, we adopted a convenience sampling strategy~\cite{hair2007_convenience_sampling_def,baltes2022_convenience_sampling_SE}, which is commonly employed in SE research when targeting specialized populations. This non-probabilistic approach relies on the availability and accessibility of participants, and although it limits the generalizability of findings, it is particularly effective when seeking input from domain-specific experts.

We targeted researchers with experience in applying LLMs to software repository mining tasks. Specifically, our inclusion criteria required participants to (i) have conducted or supervised empirical studies involving LLMs for repository analysis, and (ii) be currently active in software engineering research. These criteria ensured that participants could provide informed and contextually grounded responses. To reach this audience, we employed a three-front recruitment strategy:
\begin{itemize}
    \item We disseminated a call for participation through professional social media channels, primarily \textsl{LinkedIn}\footnote{\url{https://www.linkedin.com/feed/update/urn:li:activity:7326243068523769856/}}, targeting academic and practitioner communities interested in LLMs and mining software repositories.
    \item We directly invited researchers from our professional networks, focusing on groups known to be actively engaged in LLM-based software engineering research.
    \item Finally, we contacted the corresponding authors of the primary studies identified during our rapid review (see Section~\ref{sec:review}).
\end{itemize}

Participation in the questionnaire survey was voluntary and anonymous. Respondents were required to provide informed consent before beginning the questionnaire, and no personally identifiable information was collected. The questionnaire survey was hosted on Google Forms and remained open for a period of two months. To ensure data reliability, the survey included an attention check question, and only responses from participants who reported direct experience with LLM-based repository mining were retained for analysis.

Starting from a candidate pool of \textbf{143} researchers identified through our recruitment strategies, we distributed the questionnaire survey using an open invitation strategy, targeting individuals with relevant expertise in LLM-based repository mining. Our goal was not to achieve statistical representativeness, but rather to gather diverse and informed perspectives from researchers directly engaged in this emerging area. Responses were collected iteratively until no substantially new themes emerged across open-ended responses, a principle aligned with theoretical saturation~\cite{saunders2018saturation}, where additional data no longer yields new insights.

In total, \textbf{37} valid responses were collected and retained for analysis. Among the 37 participants, only one participant answered the question about attention-check (see next subsection) incorrectly and was removed from the survey. In total, \textbf{22} respondents confirmed their experience with the application of LLMs in repository mining and have passed the eligibility check embedded in the questionnaire survey. 

To provide context for our questionnaire survey findings, participants held a variety of research positions, with the majority being PhD students (9 participants, 40.91\%). The rest of the sample included Associate Professors (4 participants, 18.18\%), Assistant Professors or Lecturers (3, 13.64\%), and Full Professors (2, 9.09\%). Other roles were represented by single participants each (4.55\%), including Academic Researchers (non-faculty), Postdoctoral Researchers, Industry Researchers, and one respondent who identified as both a PhD Student and an Academic Researcher (non-faculty).
Participants reported varying levels of experience in Empirical Software Engineering. The most common range was 3–5 years (9 participants), followed by 6–9 years (5 participants). A smaller number of respondents reported less than 2 years (4 participants) or more than 10 years (4 participants) of experience in the field. When asked about their prior experience conducting LLM-based studies in MSR, most participants reported having conducted 2–3 studies (11 participants), followed by those who had conducted a single study (8 participants). A smaller group (3 participants) indicated having conducted more than 3 studies.

\subsubsection{Questionnaire Survey Data Extraction and Analysis}
\label{sec:survey_analysis}

After collecting the questionnaire survey responses, we performed a data cleaning process to ensure the reliability and validity of the dataset. The first author manually reviewed each submission to assess completeness, internal consistency, and attentiveness. This step is essential in survey studies, especially in online environment, where inattentive or low-effort responses can compromise data quality \cite{kitchenham2008_PersonalOpinionSurveys,stol2018grounded}. Additionally, we included an attention-check question in the questionnaire survey to identify disengaged participants. The flagged responses were jointly reviewed by the first and second authors, and final inclusion decisions were made by consensus, as recommended in qualitative questionnaire survey reliability procedures \cite{cruzes2011recommended}. Once the dataset was validated, we analyzed the responses aligned with our research questions. Responses to closed-ended questions were analyzed using descriptive statistics, including frequency distributions and visual aggregations, to identify patterns across participant practices, tool usage, prompt configurations, and validation strategies. This analysis enabled a structured understanding of how LLMs are currently being used in MSR. Open-ended responses were reviewed collaboratively by the first and second authors using qualitative content analysis~\cite{krippendorff2018content}. Through iterative reading and topic-driven grouping, we summarized responses into thematically consistent categories. This method supports the systematic description of unstructured textual data and is well-established in empirical software engineering for interpreting practitioner feedback~\cite{cruzes2011recommended, runeson2009guidelines}. The results from these two complementary analyses were then mapped to our research questions.

\textbf{RQ\textsubscript{1}} was addressed through questions in Section 2. These questions explored the types of prompts used, the specific LLMs and repositories involved, the activities performed, parameter settings, validation methods, automation practices, and the metrics employed.
\textbf{RQ\textsubscript{2}}, which focuses on the threats that researchers encounter when using LLMs for repository mining, we gathered insights through questions in Sections 3 and 4 of the survey. These questions asked participants to describe their most significant threats and the strategies they employed to overcome them. 

%% file: Section/Results.tex
\section{Analysis of the Results}
\label{sec:results}
This section presents the synthesized findings from our mixed-method study, combining a rapid review of 15 peer-reviewed articles with a questionnaire survey of 22 experienced researchers. By triangulating evidence from published literature and practitioner experience, we address our research questions on the approaches and threats in LLM-based mining of software repositories. Although all responses were collected anonymously, for readability, we assign participant identifiers (R1–R22) when quoting or referring to individual answers.

\subsubsection{Context of the Analyzed Studies: Research Topics and Platforms}
Before detailing the methodological approaches, we first characterize the context from which our evidence is drawn. Our analysis of the literature and questionnaire survey responses reveals that LLM-based MSR is being applied to a consistent set of research themes. The most predominant theme in the rapid review papers is \textbf{Code Quality, Maintenance \& Security}, which was the focus of 6 out of 15 studies, covering topics like detecting green architectural tactics \cite{de2024developers}, code smells \cite{silva2024detecting}, code-comment inconsistencies \cite{zhang2024leveraging}, self-admitted technical debt \cite{sheikhaei2024empirical}, security vulnerabilities \cite{wu2024semantic}, and notebook executability issues \cite{nguyen2025majority}. The second most common theme is \textbf{Developer Communication \& Sentiment}, addressed by five studies that analyzed emotions \cite{imran2024uncovering, kim2025exploring}, sentiment \cite{zhang2025revisiting, shafikuzzaman2024empirical}, and figurative language \cite{chatterjee2024shedding} in developer discourse. This is followed by \textbf{Repository Intelligence \& Metadata}, with three studies focusing on creating datasets \cite{jiang2024peatmoss}, analyzing software failures from news \cite{anandayuvaraj2024fail}, and building repository chatbots \cite{abedu2025repochat}. Finally, one study focused on \textbf{Issue Management}, specifically on issue report classification \cite{colavito2025benchmarking}.

This distribution of research topics is strongly mirrored in the activities reported by our survey respondents. A majority of practitioners (\textbf{13 out of 22}) focus on topics related to \textbf{Code Quality, Maintenance, and Security}, with respondents listing activities such as \quoted{Code smell detection, technical debt}, \quoted{MSR for security}, \quoted{defect prediction}, and \quoted{Architecture reconstruction}. \textbf{Issue Management} was also a common focus, mentioned by \textbf{seven} respondents with tasks like \quoted{issue report classification} and \quoted{bug localization}.

\begin{figure}[h]
    \centering
    \includegraphics[width=0.9\linewidth]{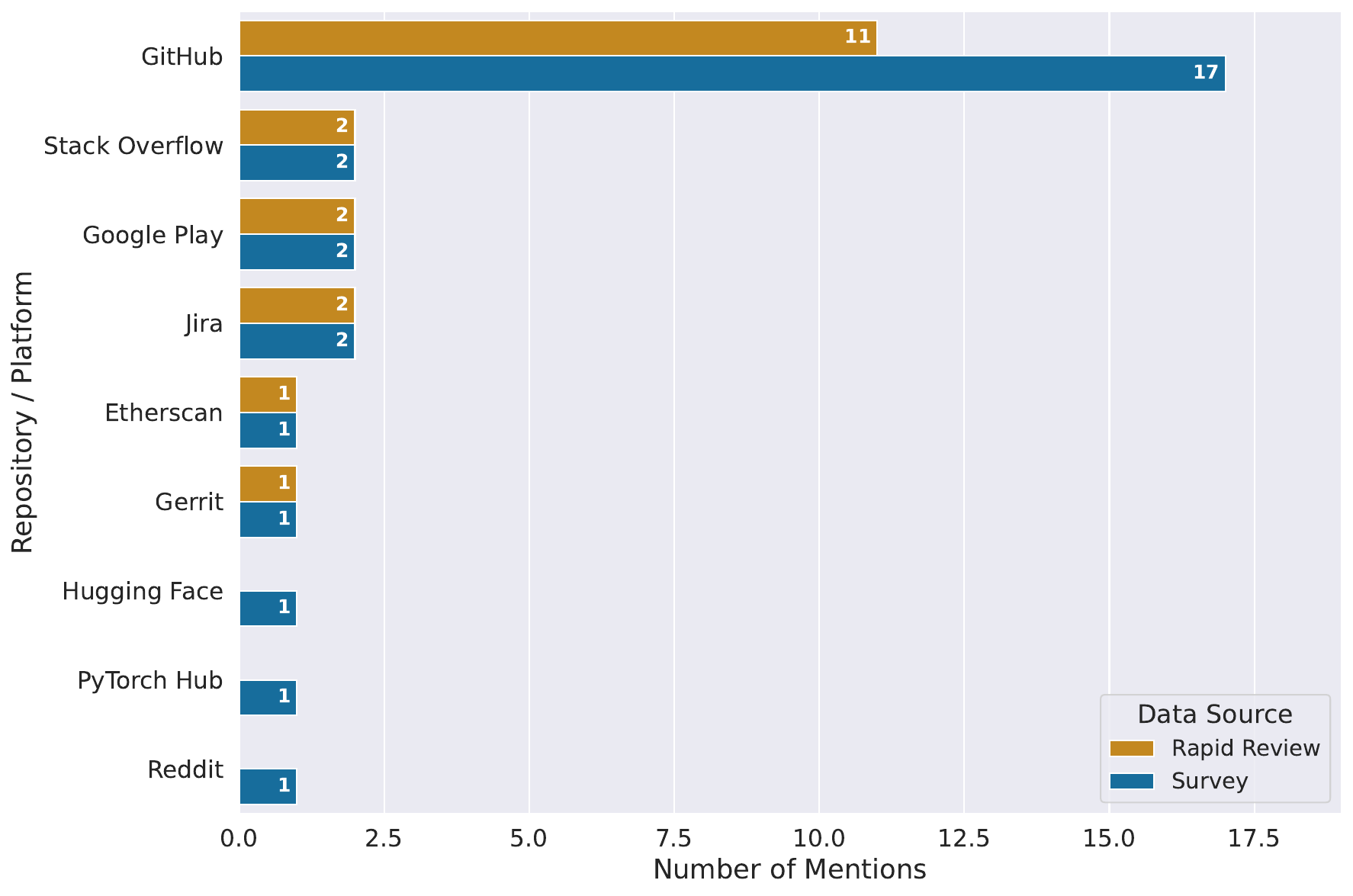}
    \caption{Frequency of software repositories and platforms analyzed in the rapid review (N=15 papers) and survey (N=22 respondents).}
    \label{fig:repo_analysis}
\end{figure}

The platforms analyzed for these studies are detailed in Figure~\ref{fig:repo_analysis}. The results show that, while some practices are consistent, others highlight emerging trends. In both the published literature and the practices reported by our questionnaire survey respondents, \textbf{GitHub} emerges as the predominant source for MSR data. However, a notable divergence appears with model-sharing platforms. \textbf{Hugging Face} and \textbf{PyTorch Hub} are frequently used by researchers in our questionnaire survey, reflecting a growing interest in analyzing ML models and their ecosystems, but they were not mentioned as primary data sources in the reviewed literature. Platforms for developer communication and issue tracking, such as \textbf{Stack Overflow} and \textbf{Jira}, appear with low frequency in both data sources. Finally, other specialized sources like \textbf{Google Play}, \textbf{Gerrit}, and \textbf{Etherscan} were used exclusively in the reviewed literature for task-specific analyses.

\subsection{RQ$_1$: What approaches are employed when using large language models for repository mining?}



The first research question aimed to investigate the methodological approaches adopted by researchers when using LLMs in MSR studies. To further improve the clarity and interpretability of our findings for \textbf{RQ$_1$}, we organized the identified practices in stages of the LLM-based research process. This staging builds upon our earlier conceptualization, which initially included four phases of PRIMES \cite{de2025framework}, as described in Section \ref{sec:primes}. However, it required refinement to better align with empirical evidence from the literature and questionnaire survey data.
The updated structure introduces two entirely new stages and refines three existing ones. \squoted{Stage 0: Strategic Planning \& Preparation} is newly introduced to explicitly account for preliminary tasks, such as defining study objectives, selecting methodological guidelines, curating datasets, and integrating automation and tooling for scalable validation workflows. These activities were previously implicit or considered in other stages but emerged as distinct and recurrent decisions in both the reviewed studies and practitioner reports.
\textsl{\quoted{Stage 1: Creation of Prompts for Piloting}} remains conceptually consistent with the original framework.
In \textsl{\quoted{Stage 2: Prompt Validation}}, we refined the title to better reflect the process of including the human in the loop.
\textsl{\quoted{Stage 3: Large Language Models Setup}} is a newly introduced stage, reflecting the configuration of LLMs, consideration of managing ensemble strategies, documenting LLM parameters, and ensuring reproducibility. 
\textsl{\quoted{Stage 4: LLM Comparison}} updates an earlier stage of PRIMES and focuses on the comparative evaluation of candidate models to select the most suitable one for the main study.
Finally, \textsl{\quoted{Stage 5: Execution, Validation \& Synthesis}} is a new, comprehensive stage covering the large-scale execution on the full dataset, the subsequent validation of outputs through error correction and post-processing, the final analysis of results, and the crucial practice of publishing replication packages.
Table~\ref{tab:approaches_summary} summarizes this structure, mapping each stage to the methodological approaches identified in our study. Although presented sequentially for clarity, the overall process is highly iterative in practice, with researchers frequently revisiting earlier stages as new insights or failures emerge during execution.

\begin{table}[ht]
\centering
\small
\caption{Summary of the stages and methodological approaches of LLM4 MSR research. Categories marked with (*) are newly introduced in this version; those marked with (\dag) represent revisions of categories from the original PRIMES framework~\cite{de2025framework}.}
\label{tab:approaches_summary}
\resizebox{\textwidth}{!}{%
\rowcolors{1}{gray!15}{white}
\begin{tabular}{|p{3cm}|p{4.5cm}|l|}
\rowcolor{black}
\hline
\textcolor{white}{\textbf{Stage}} & \textcolor{white}{\textbf{Description}} & \textcolor{white}{\textbf{Identified Approaches}} \\
\hline
\textbf{Stage 0: Strategic Planning \& Preparation *} & Defining research goals, considering established standards, automating and tracking the workflow, and preparing the dataset before model interaction. & \faCogs~ \textbf{A1: Use of Existing Guidelines} \\
\cline{3-3}
& & \faCogs~ \textbf{A2: Goal and Task Definition} \\
\cline{3-3}
& & \faCogs~ \textbf{A3: Dataset Preparation and Curation} \\
\cline{3-3}
& & \faCogs~ \textbf{A4: Automation \& Track the Process} \\
\hline

\textbf{Stage 1: Creation of Prompts for Piloting} & Designing the initial prompt structure, specifying output formats, and selecting an LLM to perform pilot testing. & \faCogs~ \textbf{A5: Selecting Prompting Strategy } \\ 
\cline{3-3}
& & \faCogs~ \textbf{A6: Select Large Language Models} \\\hline

\textbf{Stage 2: Prompt Validation \dag} & Evaluating and refining the prompt using pilot runs, agreement metrics, and annotated samples. & \faCogs~ \textbf{A7: Iterative Prompt Refinement and Pilot Testing} \\ 
\cline{3-3}
& & \faCogs~ \textbf{A8: Evaluation Metrics} \\
\hline

\textbf{Stage 3: Large Language Models Setup *} & Selecting, configuring, documenting, comparing multiple LLMs used in the study, and considering the adoption of ensemble strategies. & \faCogs~ \textbf{A9: LLM Reporting Configuration} \\
\cline{3-3}
& & \faCogs~ \textbf{A10: Consider LLMs Ensemble} \\\hline

\textbf{Stage 4: LLM Comparison \dag} & Benchmarking LLMs using validated prompts to select the optimal LLM based on predefined metrics on an oracle. & \faCogs~ \textbf{A11: Output Validation Methods} \\
\cline{3-3}
& & \faCogs~ \textbf{A12: Benchmarking and Comparative Analysis} \\\hline

\textbf{Stage 5: Execution, Validation \& Synthesis \dag} & Executing the chosen LLM at scale, validating outputs through post-processing and error correction, synthesizing final results, and publishing a replication package. & \faCogs~ \textbf{A13: Large-Scale Execution} \\
\cline{3-3}
& & \faCogs~ \textbf{A14: Result Analysis and Synthesis}\\
\cline{3-3}
& & \faCogs~ \textbf{A15: Open-Source Replication Packages} \\\hline

\end{tabular}
}
\end{table}

\paragraph{\textbf{\underline{Stage 0: Strategic Planning \& Preparation *}}} This phase lays the foundations for the study by aligning the research objectives, methodological choices, and resource organization. It ensures that subsequent activities are based on a coherent framework that promotes consistency, scalability, and reproducibility.

\begin{description}[leftmargin=0.3cm]

\item \faCogs~\textbf{A1: Use of Existing Guidelines.} Adhering to established methodological guidelines is a fundamental approach to ensuring rigor in LLM-based MSR and, more broadly, in empirical software engineering research. Our rapid review suggests that adherence to formal guidelines is not yet a common practice in published work. Only two studies (out of 15) explicitly cited external guidelines, both referring to best practices from OpenAI \cite{abedu2025repochat, shafikuzzaman2024empirical}. One study~\cite{imran2024uncovering} explicitly derived its prompt design from existing studies in the literature. The majority of papers (13) developed their methodologies without referencing established standards.

In the questionnaire survey, we asked participants whether they follow any guidelines or best practices when using LLMs. Out of 22 responses, the largest group, comprising 10 participants (45.5\%), stated they were unaware of such guidelines or responded with \squoted{NA}, highlighting a gap in awareness or accessibility of structured guidance. Only 2 respondents (9.1\%) reported following established frameworks or formal guidelines, including references to the LLM4SE guidelines \cite{wagner2024towards} and the PRIMES framework \cite{de2025framework}. A larger portion, 7 participants (31.8\%), reported adherence to informal best practices, such as those recommended by OpenAI or Google, or general principles of prompt engineering and research ethics. Finally, 3 respondents (13.6\%) described partial or exploratory adoption, indicating either inspiration from related work or intentions to follow formal practices in future studies.

While a minority of researchers follow formal frameworks or guidelines, a large group adopts informal strategies, and many researchers continue to operate without structured methodological support. This underscores the need for better dissemination and adoption of industry-specific LLM guidelines in empirical software engineering.

\smallskip
\item \faCogs~\textbf{A2: Goal and Task Definition.} Another essential methodological step, preceding any technical implementation, is the strategic definition of research objectives. This initial planning phase, which directly guides prompt creation and the overall study design, was consistently identified as the starting point in our data. Survey respondents explicitly listed activities like \textit{\quoted{Define the goal of the analysis}} (R1) and \textit{\quoted{Define Task}} (R13) as the first actions in their workflows. This approach ensures that all subsequent methodological choices, from data curation to model selection, are aligned with a clear and well-defined research goal.
\smallskip

In line with these goals, studies typically begin by identifying and collecting a diverse range of software artifacts that match the targeted research questions. Published work analyze code-level data, such as entire code repositories to identify architectural tactics \cite{de2024developers}, individual classes to detect code smells \cite{silva2024detecting}, smart contract source code for security vulnerabilities \cite{wu2024semantic}, and code comments to find self-admitted technical debt \cite{sheikhaei2024empirical} or inconsistencies \cite{zhang2024leveraging}. Other works focus on developer communication and process artifacts, including issue reports for classification \cite{colavito2025benchmarking} and discussions in commits or pull requests for sentiment and emotion analysis \cite{imran2024uncovering, zhang2025revisiting, chatterjee2024shedding}. The scope extends even to non-traditional artifacts like computational notebooks for executability analysis \cite{nguyen2025majority}, model cards from platforms like Hugging Face for metadata extraction \cite{jiang2024peatmoss}, and external sources like online news articles to analyze software failures \cite{anandayuvaraj2024fail}.
\end{description}

\begin{description}[leftmargin=0.3cm]
\item \faCogs~\textbf{A3: Dataset Preparation and Curation.} Following the definition of research objectives, a critical preliminary step is the preparation and curation of the dataset. Our review of the literature shows that studies consistently perform a data collection and preprocessing phase before engaging with any LLM. 

This involves identifying and gathering relevant software artifacts. The scope of these artifacts is broad, ranging from granular code-level data to project-wide communications and even external sources. For instance, studies have focused on individual code comments to identify Self-Admitted Technical Debt \cite{sheikhaei2024empirical}, specific code snippets containing known smells \cite{silva2024detecting}, and function-level code with its accompanying comments to detect inconsistencies \cite{zhang2024leveraging}. Other research analyzes artifacts at a larger scale, such as entire code repositories to find green architectural tactics \cite{de2024developers} or smart contracts to detect security vulnerabilities \cite{wu2024semantic}. A significant portion of studies also targets developer communication, analyzing issue reports from bug tracking systems \cite{colavito2025benchmarking}, pull request discussions \cite{zhang2025revisiting}, and general developer messages to understand emotions \cite{imran2024uncovering} or figurative language \cite{chatterjee2024shedding}. The spectrum of artifacts also includes more specialized data, such as computational notebooks to assess their executability \cite{nguyen2025majority}, model cards from platforms like Hugging Face \cite{jiang2024peatmoss}, and even news articles reporting on software failures \cite{anandayuvaraj2024fail}. Beyond simple collection, this phase often includes significant data curation efforts, such as cleaning noisy text, filtering for relevant samples, and preprocessing the data (e.g., through truncation or dataset balancing) to create a high-quality input corpus suitable for the LLM \cite{sheikhaei2024empirical}.

Beyond simple collection, this phase often includes significant data curation efforts, such as cleaning noisy text, filtering for relevant samples, and preprocessing the data (e.g., through truncation or dataset balancing) to create a high-quality input corpus suitable for the LLM \cite{sheikhaei2024empirical}.

This practice is strongly confirmed by our questionnaire survey respondents, who identify it as important part of their workflow. Researchers explicitly listed activities such as \textit{\quoted{data collection, pre-processing}} (R17) and \textit{\quoted{Input collection}} (R2) as essential steps that precede prompt engineering. This stage ensures that the data fed to the LLM is clean, relevant, and structured, which is a prerequisite for designing effective prompts and obtaining reliable results.

\smallskip
\item \faCogs~\textbf{A4: Automation \& Track the Process.} To manage the complexity of the entire workflow, from request to validation, researchers create a pipeline using customized automation tools. This practice is consistently observed in the literature and our questionnaire survey. Many studies report the development of bespoke systems, such as FAIL \cite{anandayuvaraj2024fail}, PonziSleuth \cite{wu2024semantic}, and RustT4 \cite{zhang2024leveraging}, or novel mining mechanisms \cite{de2024developers, jiang2024peatmoss} to orchestrate the analysis. This is corroborated by our survey respondents, who describe \textit{\quoted{Python scripts, and I create an automated pipeline that runs all the phases}}(R2) to drive the entire research workflow, the same mechanisms are used in literature \cite{de2024developers,silva2024detecting}. A primary target for automation is the handling of inconsistent LLM outputs. Researchers frequently build scripts to validate and parse responses, often employing fallback mechanisms. For example, one practitioner detailed a common strategy: \textit{\quoted{The output validation and extraction are performed by parsing the output automatically (if it has the requested structure); otherwise, I use regular expressions to find an answer}} (R15). This approach is documented in the literature, where regular expressions are used to reliably extract data from malformed model outputs \cite{abedu2025repochat}. Beyond parsing, some researchers leverage automation for the creative process of prompt engineering itself, including using LLMs for \textit{\quoted{automatic prompt engineering}} (R7) or applying \textit{\quoted{meta prompting when generating prompts}} (R3) to improve their quality iteratively.

This custom tooling often integrates with other libraries and analysis techniques. For instance, questionnaire survey respondents report combining their scripts with standard repository mining libraries, such as \textit{\quoted{PyDriller or simple GraphQL queries}} (R3). Similarly, our literature describes pipelines that integrate LLMs with static analysis \cite{wu2024semantic, zhang2024leveraging} or automated execution frameworks, such as Papermill \cite{nguyen2025majority}. This investment in custom automation is critical not only for managing complexity but also for enhancing scientific rigor. By codifying the entire experimental workflow, these automated pipelines effectively act as an executable specification of the research method. They ensure reproducibility by capturing the exact versions of the dataset, the final \textbf{prompts}, the complete LLM configuration (including model version, temperature, and seed), and the logic for output parsing and validation. This makes the entire research process more transparent and replicable by researchers.
\end{description}

\paragraph{\textbf{\underline{Stage 1: Creation of Prompts for Piloting}}} This phase defines the prompt structures, specifying output formats, and selecting initial models for piloting. It enables early iterations to align the model’s behavior with the research goals and constraints.

\begin{description}[leftmargin=0.3cm]
\smallskip
\item \faCogs~\textbf{A5: Selecting Prompting Strategies.} The design of prompts is one of the most important methodological aspects. In the reviewed literature, several distinct strategies have been employed to train LLMs. These strategies are also highlighted by Hou et al. \cite{hou2024large} as follows:
\begin{itemize}
    \item \textbf{Zero-shot prompting}, where the model is expected to perform a task without any explicit training or examples, relying solely on the provided instructions \cite{radford2019language}. This was the most common pattern, used in 12 of the 15 reviewed studies \cite{imran2024uncovering, silva2024detecting, colavito2025benchmarking, nguyen2025majority, zhang2024leveraging, wu2024semantic, anandayuvaraj2024fail, sheikhaei2024empirical, zhang2025revisiting, chatterjee2024shedding, shafikuzzaman2024empirical, kim2025exploring}, underscoring a reliance on the out-of-the-box capabilities of powerful models.

    \item \textbf{Few-shot prompting}, which involves providing a limited number of examples within the prompt to help the model learn the task and generalize to similar cases \cite{brown2020language}. This was the second most common technique, used in six studies to improve performance with in-context examples \cite{colavito2025benchmarking, sheikhaei2024empirical, zhang2025revisiting, chatterjee2024shedding, shafikuzzaman2024empirical, kim2025exploring}.

    \item \textbf{Chain-of-Thought (CoT) prompting}, an advanced technique that guides the model to break down a problem into intermediate reasoning steps, enhancing its ability to tackle complex tasks \cite{wei2022chain}. This emerging strategy was identified in three reviewed studies \cite{wu2024semantic, anandayuvaraj2024fail, colavito2025benchmarking}.
\end{itemize}

Beyond selecting a single prompting strategy, an identified approach in our review is the \textbf{use of multiple prompts} to achieve a research goal. A common pattern is \textbf{Prompt Chaining} \footnote{Chain complex prompts for stronger performance \url{https://docs.anthropic.com/en/docs/build-with-claude/prompt-engineering/chain-prompts}}, where a complex task is decomposed into a sequence of simpler prompts. For instance, one prompt might first classify an artifact (e.g., an emotion), and a second prompt then uses that output to perform a more detailed analysis (e.g., extract the cause of the emotion) \cite{imran2024uncovering,wu2024semantic}.

Our survey data shows similar findings.  \textbf{Zero-shot prompting} remains the most widespread technique (used by 20 of 22 respondents), often serving as a baseline. However, practitioners frequently adopt more complex strategies to improve reliability. For instance, the concept of prompt chaining is mirrored in practice, with one researcher describing how they \textit{\quoted{split the prompts into multiple prompts, making the LLM go into the repository gradually}} (R5) to build context incrementally. Furthermore, to ensure predictable and parsable outputs, researchers are increasingly adopting \textbf{Structured Chain-of-Thought (SCoT) Prompting}~\cite{li2025structured}, which was used by 18 of 22 respondents. The motivation is to gain control over the output, as one respondent aims to \textit{\quoted{tailor the prompts with detailed descriptions to generate most consistent (at least correct) results}} (R11). This is followed by \textbf{Few-shot prompting} (17 of 22) and standard \textbf{Chain-of-Thought prompting} (13 of 22), where researchers guide the model by asking it to, for example, \textit{\quoted{explain the classes involved and their interaction}} (R5) before providing a final answer. This suggests that while zero-shot is a common starting point, researchers in MSR actively employ a range of sophisticated prompting methods to control model behavior and ensure reliable results.

\smallskip
\item \faCogs~\textbf{A6: Select Large Language Models.}  With a prepared dataset, the next approach is the selection of the LLM itself. A wide array of LLMs, both proprietary and open-source, were utilized in the literature.
\begin{itemize}
    \item \textbf{Proprietary Models:} The landscape is dominated by OpenAI's GPT series. \textbf{GPT-4} and its variants (e.g., GPT-4o, GPT-4-turbo) were used in nine studies \cite{imran2024uncovering, de2024developers, jiang2024peatmoss, abedu2025repochat, zhang2024leveraging, anandayuvaraj2024fail, zhang2025revisiting, shafikuzzaman2024empirical, kim2025exploring}, and \textbf{GPT-3.5-turbo} was used in seven studies \cite{imran2024uncovering, silva2024detecting, colavito2025benchmarking, jiang2024peatmoss, wu2024semantic, anandayuvaraj2024fail, shafikuzzaman2024empirical}. This highlights a reliance on high-performing, commercially available models. Other models like Anthropic's \textbf{Claude} series \cite{de2024developers} and Google's \textbf{Gemini} family \cite{kim2025exploring} also appeared, suggesting diversification.
    
    \item \textbf{Open-Source Models:} There is a growing adoption of open-source alternatives. Smaller, encoder-only models (sLLMs), such as \textbf{BERT}, \textbf{RoBERTa}, and \textbf{CodeBERT}, were frequently used as baselines or for fine-tuning in four studies \cite{sheikhaei2024empirical, zhang2025revisiting, chatterjee2024shedding, shafikuzzaman2024empirical}. Larger, decoder-only open-source models like \textbf{LLaMA} \cite{nguyen2025majority, wu2024semantic, zhang2025revisiting}, \textbf{Flan-T5} \cite{imran2024uncovering, sheikhaei2024empirical}, \textbf{Mistral} \cite{wu2024semantic, kim2025exploring}, \textbf{Vicuna} \cite{zhang2025revisiting}, and \textbf{Qwen} \cite{kim2025exploring} were also employed, indicating a trend towards leveraging models that offer greater transparency and customization.
\end{itemize}

This trend of using a diverse set of models is directly reflected in the practices of our survey respondents. Confirming the dominance of proprietary solutions, GPT-4 was the most-used LLM, employed by 68.2\% of researchers (15 out of 22). Close behind, LLaMA 3 was reported by 11 respondents (50\%), highlighting a strong adoption of open-weight models, particularly Meta’s LLaMA series. This strong adoption is likely driven by several factors that directly address key threats identified in our study. Open-weight models provide a practical solution to the high financial costs and scalability issues associated with proprietary APIs, two of the most frequently reported threats by our survey respondents. Furthermore, running models on local or controlled infrastructure provides researchers with greater control over the experimental environment, thereby enhancing reproducibility by mitigating the model drift that is common in commercial APIs. The ability to fine-tune these models on domain-specific data, combined with the competitive performance of recent releases, makes them a compelling alternative for research.

GPT-3 remains widely used, cited by nine participants (40.9\%), likely due to legacy system integration or accessibility. Claude 3 (27.3\%) and Gemini 2.5 Pro (22.7\%) were also frequently selected, reflecting the growing competitiveness of Anthropic and Google’s offerings in the LLM ecosystem.
Other models showed more niche adoption:
\begin{itemize}
    \item Qwen and Gemma were each mentioned three times;
    \item Mixtral, Deepseek v1, and DeepSeek V3 mentioned two times;
    \item while Claude 3.5 Sonnet, Gemini 1.5 Pro, Gemini 1.5 Flash, Gemini 2.0 Flash, GPT4-o, CodeQwen, GPT-o4-mini, Deepseek-coder, DeepSeek, codegeex, codellama, starling-lm, phi, starcoder, mistral, and zephyr were each mentioned one time.
\end{itemize}
Finally, custom models, including those trained internally, were reported by three participants, indicating that while most researchers rely on pre-trained foundation models, a subset is investing in domain-specific or fine-tuned solutions tailored to their specific needs.

These results underscore a dual trend: the continued dominance of powerful general-purpose models like GPT-4, and a diversification toward open-weight and customizable alternatives.

Most importantly, the process of choosing among these models is often empirical, as articulated by one researcher whose workflow includes an explicit step to \textit{\quoted{compare the final prompt on multiple LLMs, and select the final LLM}} (R1). This phenomenon is also highlighted in literature \cite{de2024developers,silva2024detecting,colavito2025benchmarking,jiang2024peatmoss,wu2024semantic}. This indicates a shared landscape where researchers leverage commercial models and open-source alternatives for transparency, often making their final selection based on comparative evaluation.
\end{description}

\paragraph{\textbf{\underline{Stage 2: Prompt Validation \dag}}} This phase focuses on evaluating and refining the prompt through small-scale tests. It introduces human-in-the-loop validation and agreement measures to ensure the prompt produces reliable and interpretable outputs.
\begin{description}
\smallskip
\item \faCogs~\textbf{A7: Iterative Prompt Refinement and Pilot Testing.} A crucial approach that bridges initial prompt design and full-scale analysis is the practice of iterative refinement through pilot testing. Rather than immediately applying a designed prompt to an entire dataset, \textbf{researchers execute it on a small but representative sample with a single LLM}. This pilot sample often serves as a "gold standard" or "oracle" for validation. For instance, studies in our review created validation sets ranging from a few dozen instances, such as 31 software failure incidents \cite{anandayuvaraj2024fail} or 50 model cards selected via stratified sampling \cite{jiang2024peatmoss}, to a set of 10 diverse projects \cite{de2024developers}. This hands-on, experimental loop is clearly described by survey respondents, with one detailing their process as a cycle of \textit{\quoted{Define Task... Experiment with Prompts... Redefine Prompt (go back to step 2, repeat)}} (13) and another characterizing it as a phase of \textit{\quoted{trial and errors on a subset}} (7) while other refined the prompt asking directly to the LLMs used \textit{\quoted{I've started with a base prompt that I improved by asking ChatGPT directly}} (R4) or \textit{\quoted{Prompts were sometimes refined leveraging LLMs}} (R16).

This practice is consistently reported in the literature, where researchers conduct pilot sessions to iteratively refine the wording, structure, and examples of prompts based on observed LLM behavior and output quality \cite{de2024developers, silva2024detecting, jiang2024peatmoss}. The core of this approach involves manually inspecting the pilot results to identify issues like hallucinations, incorrect formatting, or misinterpretations, and then adjusting the prompt accordingly. A detailed workflow provided by one respondent includes: \textit{\quoted{Initial prompt drafting; Validation of the Draft’s Results; Prompt Refinement; Prompt Execution on a small subset}} (R3).

While refining prompts using a single LLM may introduce bias when results are later compared across multiple models, this remains a practical and effective strategy for aligning prompt behavior with the research goals. This iterative validation ensures that the prompt is robust and reliable before being used for large-scale execution, thereby saving time, reducing costs, and minimizing the risk of collecting low-quality data.

\smallskip
\item \faCogs~\textbf{A8: Evaluation Metrics.} To quantify the results of the validation process, researchers employ a well-established set of metrics, largely drawn from traditional classification, statistical comparison, and natural language generation tasks. The activity is reported as \textit{\quoted{Prompt Execution on a small subset, Results evaluation, Refinement(if necessary)}} (R2). The reviewed literature consistently features standard classification metrics such as \textbf{Accuracy}, \textbf{Precision}, \textbf{Recall}, and \textbf{F1-score} \cite{silva2024detecting, sheikhaei2024empirical, zhang2024leveraging}. These are often complemented by statistical tests to compare model or prompt performance, such as \textbf{McNemar’s test}, often reported with \textbf{Odds Ratio} and \textbf{Absolute Risk Difference} \cite{silva2024detecting, shafikuzzaman2024empirical}, or the \textbf{Wilcoxon signed-rank test} \cite{zhang2025revisiting}.

For tasks involving text generation or assessing semantic similarity, a different set of metrics is used. These include \textbf{BLEU} for evaluating generated text \cite{imran2024uncovering}, \textbf{cosine similarity} for comparing sentence embeddings \cite{chatterjee2024shedding}, and the Area Under the Curve (\textbf{AUC}) for evaluating classifier performance \cite{zhang2025revisiting}. Inter-rater reliability for manual validation is commonly measured with \textbf{Cohen's Kappa} \cite{colavito2025benchmarking, sheikhaei2024empirical}. Furthermore, some studies adopt highly task-specific metrics, such as the \textbf{executability ratio} for notebook restoration \cite{nguyen2025majority} or qualitative \textbf{taxonomy agreement} for analyzing failure reports from news \cite{anandayuvaraj2024fail}. Our survey data shows a strong consensus on the use of these core metrics. The most frequently reported metric was \textbf{Precision}, mentioned by 16 out of 22 respondents, followed closely by \textbf{F1-score} (15 respondents), \textbf{Accuracy} (14 respondents), and \textbf{Recall} (14 respondents). These were often used in combination to provide a comprehensive view of model performance. Beyond standard classification, a subset of researchers considered more resource-oriented measures. \textbf{API Cost} was reported by four respondents, often to compare the efficiency of different strategies, while \textbf{inference time} and the \textbf{number of tokens used} were also cited. Other specific metrics mentioned by participants included \textit{BLEU}, \textit{Mojo distance}, \textit{Exact Match}, and \textit{cosine similarity}, aligning with the diverse toolkit found in the literature.

Finally, to move beyond purely quantitative performance, researchers leverage qualitative and explanatory methods. These include \textbf{thematic analysis} for interpreting subjective outputs, as reported by our survey respondents, and the use of Explainable AI (XAI) techniques like \textbf{SHAP explanations} to understand the drivers of model predictions \cite{shafikuzzaman2024empirical}. These findings suggest a growing need to complement quantitative evaluations with interpretability and cost-awareness, especially as LLMs are applied to increasingly diverse MSR use cases.
\end{description}

\paragraph{\textbf{\underline{Stage 3: Large Language Models Setup *}}} This phase involves selecting, configuring, and documenting the LLMs used in the study. It includes decisions about model versions, parameter settings, and ensemble strategies to ensure reproducibility.

\begin{description}

\smallskip
\item \faCogs~\textbf{A9: LLM Reporting Configuration.} Once a model is selected, the next step is to define its operational parameters to ensure reproducibility and control its behavior. To enhance reproducibility, details on LLM parameter settings were often provided in the literature. For inference-based tasks, the \texttt{temperature} parameter was the most frequently configured, commonly set to zero or a low value to ensure deterministic and factual outputs \cite{colavito2025benchmarking, abedu2025repochat, anandayuvaraj2024fail, sheikhaei2024empirical, zhang2025revisiting, shafikuzzaman2024empirical}. Parameters controlling response length, such as \texttt{max\_tokens} or \texttt{max\_new\_tokens}, and overall context window size were also frequently specified to manage cost and output structure \cite{silva2024detecting, colavito2025benchmarking, jiang2024peatmoss, sheikhaei2024empirical}. For studies involving model adaptation, researchers reported fine-tuning hyperparameters, including the number of \texttt{epochs}, the \texttt{learning rate}, and the \texttt{batch size} to optimize performance on specific SE datasets \cite{sheikhaei2024empirical, zhang2025revisiting, shafikuzzaman2024empirical}. On the other hand, other studies do not define the specified parameters in the article or use default parameters \cite{de2024developers,imran2024uncovering,nguyen2025majority,chatterjee2024shedding}.

The configuration practices reported by our survey respondents align closely with these findings, particularly in terms of managing determinism and cost. The most common practice mentioned was controlling the \texttt{temperature} (nine respondents). A majority of respondents who configure parameters reported setting the temperature to a low value or zero to ensure deterministic and reproducible outputs. One researcher provided a task-dependent heuristic: using a \textit{\quoted{low temperature for classification, higher temperature for generation.}} Controlling \texttt{max\_tokens} was the second most cited strategy, used both to manage API costs and to ensure concise outputs for specific tasks. As one participant noted, for classification tasks, \textit{\quoted{putting a tailored constraint in the max token can help}} limit costs. 
Several respondents (six respondents) reported not altering any default settings. This suggests that for some researchers, default behavior is considered sufficient, or that configurability is limited by tooling or time constraints.

Beyond direct tuning, a few participants mentioned relying on guidance from literature, whitepapers, or official API documentation to inform their choices. A smaller group reported using more systematic approaches, \textit{\quoted{I use reference from the literature, otherwise I will proceed with an ablation study or a configuration experiment}} (R2) or \textit{\quoted{I use grid search to find the optimal value for a parameter}} (R15). 
Overall, these responses reveal a wide range of practices, from using default settings to a more deliberate, literature-based configuration. Although not all studies or researchers systematically adjust parameters, those who do recognize the value of improving reproducibility. 

\smallskip
\item \faCogs~\textbf{A10: Consider LLMs Ensemble.} Researchers may consider adopting an \textbf{ensemble of LLMs}, which involves combining the outputs of LLMs potentially differing in architecture, training data, or behavior, to enhance robustness and accuracy. Instead of relying on a single model, this approach leverages model diversity to mitigate individual limitations and capture a broader range of correct outputs. For extracting structured metadata from model cards, this technique proved to be advantageous ~\cite{jiang2024peatmoss}. Additionally, respondents also corroborate this approach: \textit{\quoted{sometimes multiple LLMs are needed when conducting multiple tasks at the same time}} (R10). To further improve the reliability and stability of LLM outputs, researchers employ \textbf{ensemble prompting}, where a single prompt is executed multiple times on the same input to mitigate randomness. The final result is then determined by aggregating the outputs, often through a majority vote \cite{zhang2024leveraging} or by averaging performance scores across runs \cite{wu2024semantic, colavito2025benchmarking}, especially when deterministic accuracy cannot be ensured through a single run. This practice was highlighted by a survey respondent saying:  \textit{\quoted{run experiments multiple time and use a vote criterion to determine the accuracy}} (R18). This is complemented by the use of \textbf{output filtering and parsing}. This involves developing scripts or rules to validate the structure of the LLM's response. For example, researchers use regular expressions to extract valid data from malformed outputs \cite{abedu2025repochat} or implement post-processing steps to resolve inconsistencies and normalize the data \cite{jiang2024peatmoss}. If an output fails these checks, it is often discarded. As one researcher (R15) explained, if automated parsing fails, \textit{\quoted{the output is considered null}}. Together, these techniques improve the reliability of the collected data.
\end{description}

\paragraph{\textbf{\underline{Stage 4: LLM Comparison \dag}}} This phase focuses on the systematic comparison and selection of the final LLM for the study. It involves benchmarking the performance of different candidate models using the validated prompts and evaluation metrics on a representative data sample to make an evidence-based decision.

\begin{description}

\smallskip
\item \faCogs~\textbf{A.12: Benchmarking and Comparative Analysis.} A common methodological practice is the use of benchmarking and comparative analysis to justify model selection. Rather than selecting a single model a priori, many reviewed studies evaluated multiple LLMs to identify the best-performing or most cost-effective option for their specific task. This often involved running the same prompts and dataset through different proprietary models (e.g., comparing GPT-4 against Claude 3 Haiku \cite{de2024developers}), different open-source models (e.g., LLaMA vs. Mistral \cite{wu2024semantic}), or a mix of both \cite{imran2024uncovering, kim2025exploring}. Some studies conducted extensive benchmarks, comparing over 20 different models to analyze trade-offs between performance and resource requirements \cite{colavito2025benchmarking}, while others performed ablation studies to isolate the impact of different model choices within a larger system \cite{wu2024semantic}.

This practice of multiple model selection is described by our survey respondents, whose workflows often include an explicit model evaluation phase. One researcher described this step as needing to \textit{\quoted{compare the final prompt on multiple LLMs, and select the final LLM}} (R1). This process based on performance metrics and also on practical constraints. For instance, another respondent highlighted the trade-off between cost and resources, explaining that they rely on \textit{\quoted{local setups of LLMs (Llama, Qwen)}} because larger proprietary models are often too slow or expensive to run on their available hardware (R13). The use of ablation studies, noted in the literature, is also confirmed in practice, with one researcher stating they would proceed with an \textit{\quoted{ablation study or a configuration experiment}} (R2) to justify their choices.
\end{description}

\paragraph{\textbf{\underline{Stage 5: Execution, Validation \& Synthesis \dag}}} The final stage of the framework encompasses the main data generation, validation, and synthesis activities. It includes the large-scale execution of the selected LLM on the entire dataset, ensuring the reliability of the generated data through post-processing and error correction, analyzing and interpreting the final results, and publishing replication packages to ensure transparency and reproducibility.

\begin{description}[leftmargin=0.3cm]
\smallskip
\item \faCogs~\textbf{A13: Large-Scale Execution.} Once both the prompt and the LLM have been validated and finalized, typically through pilot testing and comparative evaluations, the next step is \textbf{large-scale execution}. This approach consists of the application of the finalized experimental setup to the entire target dataset. It represents the transition from methodological adjustments to the large-scale collection of the study's raw empirical data.

This involves applying the finalized prompt to every data point, which demands significant engineering effort. This is corroborated by our survey, where a majority of respondents (14 out of 22) reported that \textbf{\quoted{high costs}} and \textbf{\quoted{scaling to large datasets}} are threats they face \squoted{Often} or \squoted{Always}. Researchers must manage concerns such as API rate limits, batch processing, hardware requirements, and overall cost \cite{de2024developers, wu2024semantic}. One respondent articulated this difficulty clearly, stating, \textit{\quoted{When scaling to large datasets, the inference time and hardware requirements are a serious problem to take into account}} (R15). The custom automation pipelines, discussed in Approach (\textbf{A14}), ensure a smooth and efficient execution in this phase. Studies in our review demonstrate this large-scale application on thousands of artifacts, such as news articles \cite{anandayuvaraj2024fail} or entire software repositories \cite{de2024developers}. Therefore, large-scale execution is not merely a final \squoted{run}, but a distinct engineering approach where researchers must balance speed, cost, and reliability to generate the full raw dataset for their study.

\smallskip
\item \faCogs~\textbf{A14: Result Analysis and Synthesis.} The final approach in the workflow is the analysis and synthesis of the validated LLM-generated data. This phase transitions from data generation to knowledge creation, where the structured outputs are interpreted to yield empirical findings that address the study's original research questions. The specific techniques used depend on the nature of the data and the research goals. For quantitative data, such as classifications or counts, this often involves descriptive statistics to analyze the frequency and distribution of categories \cite{de2024developers, silva2024detecting}. For more qualitative or unstructured outputs, researchers apply techniques like thematic analysis or clustering (e.g., using DBSCAN) to identify emergent patterns and group related concepts \cite{imran2024uncovering}. This final approach is a consistent part of the process described by survey respondents, who listed activities like \textit{\quoted{result analysis}} (R17) and \textit{\quoted{qualitative analysis through thematic analysis}} (R7) as part of their workflow. It is in this phase that the raw, machine-generated data is transformed into human-interpretable insights. For example, after using an LLM to identify thousands of software failures from news articles, one study performed a final analysis to identify trends in failure recurrence and severity over time \cite{anandayuvaraj2024fail}. This highlights that the LLM is often a powerful intermediate tool for data extraction and classification, but the ultimate research contribution comes from the subsequent analysis and synthesis performed by the researcher.

\smallskip
\item \faCogs~\textbf{A15: Open-Source Replication Packages.} An important approach for ensuring transparency and reproducibility, adopted by all 15 studies in our review, is the creation and public release of a comprehensive replication package. This approach ensures that the entire research process, from data collection to analysis, is verifiable and reusable by the community.

Our analysis of these packages reveals a common set of platforms used for dissemination. \textbf{GitHub} is the most prevalent choice, used in 9 of the 15 studies for hosting source code and versioned artifacts \cite{imran2024uncovering,jiang2024peatmoss,nguyen2025majority,zhang2024leveraging,wu2024semantic,sheikhaei2024empirical,zhang2025revisiting,chatterjee2024shedding,kim2025exploring}. This is often supplemented by long-term archival services, such as \textbf{Zenodo} (3 studies) \cite{silva2024detecting,abedu2025repochat,anandayuvaraj2024fail} and \textbf{Figshare} (3 studies) \cite{de2024developers,colavito2025benchmarking,shafikuzzaman2024empirical}, to ensure the persistent availability of the data.

The contents of these packages consistently feature artifacts for replication. For instance, the \textbf{source code} for the analysis is provided in 14 of the 15 packages, while the \textbf{datasets} used (including raw data, oracles, and final results) are also included in 14 studies. The provided code ranges from data processing scripts and Jupyter notebooks~\cite{sheikhaei2024empirical} to complete, installable tools that replicate the study's functionality~\cite{wu2024semantic, zhang2024leveraging}. Most important for this line of research, 6 of the 15 studies explicitly include the exact \textbf{LLM prompts} or prompt templates,  component for reproducing generation-based results~\cite{de2024developers, anandayuvaraj2024fail, colavito2025benchmarking}. This approach of providing comprehensive, publicly accessible artifacts is fundamental to the verifiability of LLM-based MSR, allowing other researchers to inspect the methodology, validate the findings, and build upon the work.
\end{description}

\definitionbox{RQ$_1$ — Large Language Model Approaches for Empirical Rigor}{As a result of \textbf{RQ$_1$}, we identified a set of 15 methodological approaches used when applying LLMs for MSR. These include strategies for prompt design, model selection, iterative validation, and automation. These findings offer future researchers a detailed overview of the technical and methodological choices that characterize current LLM-based MSR studies. Researchers can design their studies effectively and align with emerging best practices in the field.}

\subsection{RQ$_2$: What are the current threats to validity faced by researchers when they use large language models for repository mining?}
\label{sec:results_rq1}
Our analysis reveals a consistent set of significant threats, triangulated between the published literature and the experiences of our survey respondents. 
In the following section, we discuss each threat and the strategies for mitigating them, summarized in Table \ref{tab:threats_summary}.

\begin{table}[htbp]
\centering
\footnotesize
\caption{Summary of Threats, Mitigation Strategies, and Employed Techniques.}
\label{tab:threats_summary}
\resizebox{\textwidth}{!}{%
\rowcolors{1}{gray!15}{white}
\begin{tabular}{|p{2.5cm}|p{6.5cm}|p{6.5cm}|}
\rowcolor{black}
\hline
\textcolor{white}{\textbf{Threat}} & \textcolor{white}{\textbf{Mitigation Strategy}} & \textcolor{white}{\textbf{Techniques Employed}} \\
\hline
\faExclamationCircle\  \textbf{T1: Hallucinations, Misinterpretation, \& Output Inaccuracy}
& \textbf{(M1) Grounding via Hybrid Approaches} & Static Analysis / AST Checks \cite{wu2024semantic, zhang2024leveraging} \\
\cline{3-3}
& & Knowledge Graph (KG) Querying \cite{abedu2025repochat} \\
\cline{3-3}
& & Retrieval-Augmented Generation (RAG) \cite{anandayuvaraj2024fail} \\
\cline{2-3}
& \textbf{(M2) Structured \& Multi-Step Prompting} & Chain-of-Thought (CoT) \cite{wu2024semantic, anandayuvaraj2024fail} \\
\cline{3-3}
& & Context Enrichment (Schemas, Taxonomies) \cite{abedu2025repochat, imran2024uncovering, de2024developers} \\
\cline{2-3}
& \textbf{(M3) Output Stabilization through Aggregation:} & Majority Voting / Averaging Results \cite{zhang2024leveraging, wu2024semantic, colavito2025benchmarking} \\
\cline{3-3}
& & Vote Criterion for Accuracy (R18) \\
\hline

\faExclamationCircle\  \textbf{T2: Prompt Engineering \& Sensitivity}
& \textbf{(M4) Systematic Iteration \& Oracle-Guided Refinement} & Pilot Testing \& Comparative Evaluation \cite{de2024developers, silva2024detecting, colavito2025benchmarking} \\
\cline{3-3}
& & Iterative Workflow (R3, R7, R13) \\
\cline{2-3}
& \textbf{(M5) Adoption of De-Facto \& Emerging Standards} & Modular Component-wise Evaluation \cite{kim2025exploring} \\
\cline{3-3}
& & Structured Templates \& Guidelines \cite{de2024developers, abedu2025repochat, shafikuzzaman2024empirical} \\
\cline{2-3}
& \textbf{(M6) Meta-Prompting \& Automated Assistance} & Meta-Prompting (R3, R4, R7) \cite{jiang2024peatmoss, de2024developers, silva2024detecting} \\
\hline

\faExclamationCircle\  \textbf{T3: Contextual Understanding \& Domain Specificity} 
& \textbf{(M7) Injecting Domain Knowledge} & Grounding in SE-specific Taxonomies \cite{de2024developers, imran2024uncovering} \\
\cline{3-3}
& & Fine-tuning on Domain-specific Data \cite{chatterjee2024shedding} \\
\cline{2-3}
& \textbf{(M8) Hybrid Program Analysis} & Pre-processing with Static/Taint Analysis \cite{wu2024semantic, zhang2024leveraging, jiang2024peatmoss} \\
\cline{2-3}
& \textbf{(M9) Intelligent Context Management} & Automated Code Slicing \cite{wu2024semantic} \\
\cline{3-3}
& & RAG \& Rule-based Truncation \cite{anandayuvaraj2024fail, colavito2025benchmarking} \\
\cline{3-3}
& & Intelligent Chunking/Summarization (R17) \\
\hline

\faExclamationCircle\  \textbf{T4: Scalability, Efficiency, \& Cost} 
& \textbf{(M10) Strategic LLM Selection} & Cost-Performance Analysis \cite{de2024developers, jiang2024peatmoss, sheikhaei2024empirical} \\
\cline{3-3}
& & Use of Local/Open-Source Models (R13) \\
\cline{3-3}
& & Model Quantization (R15) \\
\cline{2-3}
& \textbf{(M11) Input \& Output Token Optimization} & Input Reduction via Code Slicing \cite{wu2024semantic} \\
\cline{3-3}
& & Output Capping with \squoted{max\_tokens} (R3, R19) \\
\cline{2-3}
& \textbf{(M12) Multi-Stage Hybrid Pipelines} & Low-cost Pre-filtering (e.g., embeddings) \cite{anandayuvaraj2024fail} \\
\hline

\faExclamationCircle\  \textbf{T5: Reproducibility \& Output Stability}
& \textbf{(M13) Deterministic Parameter Configuration} & Setting \squoted{temperature = 0} \cite{colavito2025benchmarking, anandayuvaraj2024fail} (R4, R7, R17) \\
\cline{2-3}
& \textbf{(M14) Stabilizing Stochastics through Aggregation} & Averaging Results of Multiple Runs \cite{wu2024semantic, colavito2025benchmarking} \\
\cline{2-3}
& \textbf{(M15) Transparent Reporting \& Version Pinning} & Documenting Specific Model Versions \cite{anandayuvaraj2024fail, colavito2025benchmarking} \\
\hline

\faExclamationCircle\  \textbf{T6: Output Formatting \& Parsing}
& \textbf{(M16) Prompt-Based Formatting Instructions} & Requesting Strict JSON Format \cite{jiang2024peatmoss, wu2024semantic} \\
\cline{3-3}
& & Providing Few-shot Examples of Format \cite{de2024developers} \\
\cline{2-3}
& \textbf{(M17) Robust Post-Processing \& Fallback Parsing} & Fallback Parsing with Regular Expressions \cite{abedu2025repochat} (R11, R15) \\
\cline{3-3}
& & Flexible Text Normalization \cite{imran2024uncovering} \\
\cline{2-3}
& \textbf{(M18) Manual Correction or Exclucion} & Manual Fixing of Malformed Outputs \cite{silva2024detecting} \\
\cline{3-3}
& & Discarding and Reporting Loss Rate \cite{colavito2025benchmarking} \\
\hline

\faExclamationCircle\  \textbf{T7: Dataset Noise, Imbalance, \& Pre-training Leakage}
& \textbf{(M19) Input Dataset Preparation \& Cleaning} & Multi-Annotator Reconciliation \cite{imran2024uncovering} \\
\cline{3-3}
& & Manual Error Analysis to Correct Labels \cite{sheikhaei2024empirical, zhang2025revisiting} \\
\cline{2-3}
& \textbf{(M20) Bias Mitigation through Prompting \& Analysis} & Prompt-based Safeguards (e.g., persona) \cite{kim2025exploring} \\
\cline{3-3}
& & Use of Interpretability Methods (R22) \\
\cline{2-3}
& \textbf{(M21) Acknowledgment of Data Leakage} & Acknowledging Threat to Construct Validity \cite{colavito2025benchmarking, jiang2024peatmoss} \\
\hline

\faExclamationCircle\  \textbf{T8: Validation Complexity}
& \textbf{(M22) Oracle Construction for Output Evaluation:} & Multi-Annotator Workflow \& Reconciliation \cite{imran2024uncovering, zhang2024leveraging} \\
\cline{3-3}
& & Reporting Inter-Rater Reliability \cite{anandayuvaraj2024fail} \\
\cline{2-3}
& \textbf{(M23) Scalable Validation via Statistical Sampling} & Validating on a Statistically Significant Sample (R10, R16) \\
\hline

\faExclamationCircle\  \textbf{T9: Lack of Standardized Tooling \& Infrastructure} 
& \textbf{(M24) Developing Custom Automated Pipelines} & Creating custom Python scripts \& pipelines (R2) \\
\cline{3-3}
& & Fallback parsing with regular expressions (R15) \\
\cline{2-3}
& \textbf{(M25) Integrating Libraries \& APIs} & Combining MSR libraries (e.g., PyDriller) with LLM APIs (R3) \\
\cline{2-3}
& \textbf{(M26) Adopting Local Inference Engines} & Using local frameworks like \squoted{ollama} or \squoted{llama.cpp} (R9, R15) \\
\hline

\end{tabular}
}
\end{table}

\begin{description}[leftmargin=0.3cm]

\item \faExclamationCircle\  \textbf{T1: Hallucinations, Misinterpretation, \& Output Inaccuracy.} A predominant threat is the propensity of LLMs to generate factually incorrect information (i.e., hallucinations) or misinterpret the input data, an issue reported across a majority of the reviewed studies \cite{imran2024uncovering, de2024developers, silva2024detecting, jiang2024peatmoss, abedu2025repochat, nguyen2025majority, anandayuvaraj2024fail, sheikhaei2024empirical, zhang2025revisiting, chatterjee2024shedding, shafikuzzaman2024empirical, kim2025exploring}. This finding is strongly corroborated by our survey, where half of the respondents (11 out of 22) report experiencing hallucinations \squoted{Often} or \squoted{Always}, with seven encountering them \squoted{Sometimes}. In contrast, three respondents classified the problem as occurring \squoted{Rarely}, and just one had \squoted{Never} experienced it. This issue was frequently cited as the most significant barrier, with one researcher describing how the LLM \textit{\quoted{approximates the information it returns and invents mock outputs in accordance with the little information it gets}} (R5). The literature further details this problem through examples of vague or generic outputs, as well as failure to distinguish between nuanced concepts \cite{abedu2025repochat,de2024developers,silva2024detecting,nguyen2025majority}. 

\smallskip 
\noindent\textbf{Mitigation Strategies.} To prevent and reduce inaccurate outputs, researchers could employ three strategies:
\begin{itemize}
    \smallskip
    \item \textbf{(M1) Grounding via Hybrid Approaches:} A primary strategy is to ground the LLM in verifiable external data rather than relying solely on its internal knowledge. This is achieved by pairing the LLM with deterministic components, such as using \textbf{static analysis} to extract code-level facts~\cite{wu2024semantic, zhang2024leveraging, jiang2024peatmoss}. This is confirmed by survey respondents who integrate traditional MSR libraries like \textit{\quoted{PyDriller or simple GraphQL queries}} (R3) into their LLM pipelines. Other hybrid techniques include structuring information into a \textbf{Knowledge Graph (KG)} that the LLM can query~\cite{abedu2025repochat}, or adopting \textbf{Retrieval-Augmented Generation (RAG)} to supply relevant context from long documents~\cite{anandayuvaraj2024fail, jiang2024peatmoss}.
    
    \smallskip
    \item \textbf{(M2) Structured \& Multi-Step Prompting:} Prompts are engineered to enforce logical consistency. This includes using \textbf{Chain-of-Thought (CoT)} prompts that require the model to outline its reasoning steps~\cite{wu2024semantic, anandayuvaraj2024fail, colavito2025benchmarking}, a practice echoed by researchers who structure their workflows to first ask the LLM to \textit{\quoted{explain the classes involved and their interaction}} before delivering a final result (R5). Practitioners also employ \textbf{multi-step prompting}, described by one as splitting \textit{\quoted{the prompts into multiple prompts... to build... knowledge}} (R5) incrementally. Furthermore, prompts are enriched with explicit context like KG schemas~\cite{abedu2025repochat}, code context~\cite{sheikhaei2024empirical}, detailed extraction rules~\cite{zhang2024leveraging}, or pre-defined taxonomies~\cite{imran2024uncovering, de2024developers}, with the goal of providing \textit{\quoted{detailed descriptions to generate most consistent (at least correct) results}} (R11). These strategies are also presented in the Approaches (\textbf{A5, A10}).

    \smallskip
    \item \textbf{(M3) Output Stabilization through Aggregation:} To improve robustness and filter out random errors, many studies employ an aggregation strategy based on multiple executions. Specifically, this involves executing the same prompt on the same model several times and then aggregating the outputs—often through a \textbf{majority vote} on the outcomes \cite{zhang2024leveraging, sheikhaei2024empirical}, or by \textbf{averaging performance scores across the runs} \cite{wu2024semantic, colavito2025benchmarking}, to determine the most reliable final result. This practice, which focuses on stabilizing the output of a single model configuration rather than combining different models, was corroborated by one of our survey respondents, who noted to \textit{\quoted{run experiments multiple times and use a vote criterion to determine the accuracy}} (R18). These strategies are also presented in the Approaches (\textbf{A5,10}).
\end{itemize}
\end{description}

\begin{description}[leftmargin=0.3cm]
\smallskip
\item \faExclamationCircle\  \textbf{T2: Prompt Engineering \& Sensitivity.} Crafting effective and reproducible prompts remains a significant hurdle, a problem exacerbated by a wider lack of awareness of methodological guidelines and standards in the field. This threat was noted in eight studies \cite{imran2024uncovering, silva2024detecting, colavito2025benchmarking, zhang2024leveraging, sheikhaei2024empirical, zhang2025revisiting, chatterjee2024shedding, kim2025exploring} and was a clear theme in our survey. The difficulty is not merely technical but systemic: our rapid review found that only two studies cited external guidelines for their process \cite{abedu2025repochat, shafikuzzaman2024empirical}. This is reflected in our survey, where 11 of the 22 respondents do not follow any established practices, with some stating they are \textit{\quoted{Not aware of such standard guidelines}} (R16).

This absence of a shared foundation makes prompt engineering a persistent operational bottleneck, encountered \squoted{Often} or \squoted{Always} by over a quarter of our respondents (six out of 22). Researchers describe the process as \textit{\quoted{a long process of trial and error}} (R3), with one identifying their most significant threat as the \textit{\quoted{Difficulty in designing the prompt as there are no clear guidelines on how to create them and which are more effective depending on the context}} (R1). This aligns with findings from the literature where LLM performance was found to be highly sensitive to subtle changes in phrasing and structure, leading to instability or performance degradation \cite{colavito2025benchmarking, zhang2024leveraging, chatterjee2024shedding}.

\smallskip
\noindent\textbf{Mitigation Strategies.} To mitigate prompt design threats, researchers are combining experimentation with the ad-hoc adoption of emerging best practices and could employ three strategies:

\begin{itemize}
    \smallskip
    \item \textbf{(M4) Systematic Iteration \& Oracle-Guided Refinement:} The most common mitigation is to embrace what respondents describe as a \textit{\quoted{long process of trial and error}} (R3) by turning it into a systematic, iterative process. Researchers conduct pilot tests on a small subset of data to refine prompts over several cycles \cite{de2024developers, silva2024detecting}. This process follows a clear workflow of \textit{\quoted{Initial prompt drafting; Validation...; Prompt Refinement}} (R3) within an iterative loop where researchers \textit{\quoted{... and redefine prompt (go back... repeat)}} (R13). This refinement is not random; it is often guided by objective feedback from a high-quality \textbf{manual oracle} (a ground-truth dataset) or an existing dataset to make evidence-based decisions about which prompt variations perform best before large-scale deployment \cite{jiang2024peatmoss, de2024developers, silva2024detecting}. This strategy is also presented in the Approach (\textbf{A7}). However, performing this refinement using a single LLM may introduce model-specific biases, leading to prompts that are inadvertently optimized for that model’s behavior. To mitigate this risk, researchers are increasingly testing prompts across \textbf{multiple LLMs} during Stage 4, as suggested in the approach (\textbf{A12}), to ensure generalizability and reduce overfitting to a particular LLM.

    \smallskip
    \item \textbf{(M5) Adoption of De-Facto \& Emerging Standards:} In the absence of formal standards, researchers seek guidance from available sources. This includes adopting modular and structured frameworks proposed in the literature, such as deconstructing prompts into standardized components (e.g., \textit{Persona}, \textit{Task Instruction}) for systematic evaluation \cite{kim2025exploring, de2024developers}. Practitioners also turn to de-facto standards from model providers like \textit{\quoted{OpenAI and Google}} \cite{abedu2025repochat, shafikuzzaman2024empirical} or adopt nascent academic guidelines like our \textit{\quoted{PRIMES framework}} (R6). In parallel, several communities are creating prompting guides to help researchers and practitioners\footnote{An example of such resources is the Prompting Guide \url{https://www.promptingguide.ai/prompts}}. These practices align with approaches (\textbf{A1,A5}), which cover the selection of prompting strategies.

    \smallskip
    \item \textbf{(M6) Meta-Prompting \& Automated Assistance:} A complementary strategy, identified primarily through our survey, is the use of \textbf{meta-prompting}, where an LLM is employed not to perform the target task, but to help refine the prompt itself. Respondents describe this as a process to \textit{\quoted{refine the promt with an LLM}} (R1) or what one explicitly calls \textit{\quoted{meta prompting when generating prompts}} (R3). This technique turns the model into a creative assistant that can suggest rephrasings or structural improvements, a workflow some term \textit{\quoted{automatic prompt engineering}} (R7). Another researcher described their method as \textit{\quoted{asking ChatGPT multiple times for help until coming up with a solution}} (R4), effectively leveraging the LLM to accelerate the prompt engineering cycle. While this was a clear practice among several survey respondents (R1, R3, R4, R7, R19), it is not yet a widely documented mitigation strategy in the formal literature we reviewed, suggesting it is an emerging, practitioner-driven technique. This is also highlighted in the approach (\textbf{A7}).
\end{itemize}
\end{description}

\begin{description}[leftmargin=0.3cm]
\smallskip
    \item \faExclamationCircle\  \textbf{T3: Contextual Understanding \& Domain Specificity.} LLMs often struggle with the deep contextual understanding required for specialized SE tasks, a finding evident in literature focused on detecting context-heavy code smells \cite{silva2024detecting}, interpreting SE-specific metaphors \cite{chatterjee2024shedding}, or understanding complex panic conditions in code \cite{zhang2024leveraging}. This threat is closely linked to the technical limitations of current models, a point emphasized by our survey participants. Over a third of respondents (8 out of 22) encounter technical limitations like context windows \squoted{Often} or \squoted{Always}. An additional five face them \squoted{Sometimes}, while seven report them as a \squoted{Rare} occurrence, and only two claimed to have \squoted{Never} encountered such issues. One researcher identified this as their primary obstacle, explaining that the context window \textit{\quoted{makes it difficult to provide enough context for accurate reasoning, especially when the model needs to understand dependencies across files or track code evolution}} (R17). To mitigate this, practitioners report adopting strategies such as \textit{\quoted{chunking the input intelligently or summarizing or filtering code segments before sending them to the model}} (R17).

\smallskip
\noindent\textbf{Mitigation Strategies.} To overcome these limitations, researchers focus on augmenting the LLM with domain-specific knowledge and employing sophisticated techniques to manage the finite context window. Researchers could employ three strategies:
\begin{itemize}
    \smallskip
    \item \textbf{(M7) Injecting Domain Knowledge:} To fix this problem, researchers explicitly provide the LLM with domain-specific knowledge. This is often done by grounding the prompt in established \textbf{SE-specific taxonomies} and definitions, such as providing a catalog of green architectural tactics~\cite{de2024developers}, formal emotion hierarchies~\cite{imran2024uncovering}, or detailed descriptions of code smells~\cite{silva2024detecting} and technical debt categories~\cite{sheikhaei2024empirical}. A more resource-intensive approach involves \textbf{fine-tuning} a model on domain-specific data. This is confirmed as a practice by our survey respondents (R22) and is documented in the literature for tasks like interpreting SE-specific metaphors~\cite{chatterjee2024shedding}, analyzing sentiment~\cite{zhang2025revisiting}, and identifying self-admitted technical debt~\cite{sheikhaei2024empirical}. This mitigation is related to approaches \textbf{(A2, A5)}, which support task definition and prompt design.

    \smallskip
    \item \textbf{(M8) Hybrid Program Analysis:} Instead of providing raw data and expecting the LLM to understand its complex relationships, studies use a hybrid approach where traditional program analysis pre-processes the context. Techniques like \textbf{static taint analysis}, filtering based on \textbf{Abstract Syntax Trees (ASTs)}~\cite{wu2024semantic, zhang2024leveraging, jiang2024peatmoss}, and \textbf{def-use analysis}~\cite{nguyen2025majority} are used to identify critical data flows or code structures. This practice is corroborated by survey respondents who use tools like \textit{\quoted{PyDriller}} (R3) to \textit{\quoted{filter code segments before sending them to the model}} (R17). The output of this analysis is then fed to the LLM, effectively simplifying the reasoning task. This mitigation is related to approach \textbf{(A4)}.

    \smallskip
    \item \textbf{(M9) Intelligent Context Management:} To work within the LLM's finite context window, a challenge one respondent called their \textit{\quoted{most significant}} because it \textit{\quoted{makes it difficult to provide enough context for accurate reasoning}} (R17), researchers adopt several intelligent context management strategies. The literature highlights techniques like \textbf{automated code slicing}~\cite{wu2024semantic}, rule-based \textbf{truncation}~\cite{colavito2025benchmarking, sheikhaei2024empirical}, and, for large textual documents, \textbf{Retrieval-Augmented Generation (RAG)}~\cite{anandayuvaraj2024fail, jiang2024peatmoss}. These formal techniques are corroborated in practice, with the same respondent describing their solution as needing to \textit{\quoted{chunk the input intelligently or summarize or filter code segments}} (R17) before sending them to the model. These strategies are also reflected in methodological practices (\textbf{A3, A4}).

\end{itemize}
\end{description}

\begin{description}[leftmargin=0.3cm]
\smallskip
\item \faExclamationCircle\  \textbf{T4: Scalability, Efficiency, \& Cost.} The practical application of LLMs at scale introduces concerns about computational and financial costs, a theme present in several reviewed papers \cite{de2024developers, jiang2024peatmoss, wu2024semantic}. These concerns were among the most severe and frequently reported threats in our survey. The majority of researchers, 14 out of 22 (63.6\%), encounter issues with both \textbf{High Costs} and \textbf{Scaling to Large Datasets} \squoted{Often} or \squoted{Always}. For high costs, seven respondents reported the issue occurring \squoted{Sometimes} and only one \squoted{Rarely}, with none claiming to have \squoted{Never} faced it. Similarly, for scaling issues, four encountered them \squoted{Sometimes} and three \squoted{Rarely}, with just one respondent reporting it was \squoted{Never} a problem. Respondents consistently described this as a primary barrier, with one noting: \textit{\quoted{When scaling to large datasets, the inference time and the hardware requirements are a serious problem to take into account}} (R15). Another simply stated that financial cost is \textit{\quoted{the most significant factor}} (R20).

\smallskip
\noindent\textbf{Mitigation Strategies.} To make large-scale analysis feasible, researchers could employ three strategies:.
\begin{itemize}
    \smallskip
    \item \textbf{(M10) Strategic LLM Selection:} Rather than defaulting to the largest and most expensive LLM, a key mitigation is to strategically select one that balances performance and cost. Studies in the literature conduct comparative analyses of different LLMs (e.g., GPT-4 vs. Claude Haiku), explicitly choosing the most cost-effective option that meets the required accuracy threshold \cite{de2024developers, jiang2024peatmoss}. This includes recognizing that state-of-the-art models may only offer marginal performance gains for a significant increase in cost \cite{sheikhaei2024empirical}. Our survey reveals the practical side of this strategy, where researchers use local, open-source models to avoid API fees (R13) and apply techniques like \textbf{model quantization} to run models on less powerful hardware (R15).

    \smallskip
    \item \textbf{(M11) Input \& Output Token Optimization:} Since cost is tied to token count, a primary focus is on optimization. For input, literature highlights using \textbf{program analysis}, such as building an AST, to programmatically \textbf{slice code} and provide the LLM with only the minimal necessary context~\cite{wu2024semantic}. Complementing this, our survey respondents emphasize controlling the output, a practice also seen in the literature where parameters controlling the output length (e.g., \squoted{max\_new\_tokens} or \squoted{max\_tokens}) are explicitly limited~\cite{sheikhaei2024empirical}. A common cost-saving measure is to strictly limit this parameter to prevent unnecessarily long responses. Respondents confirm this practice, stating its purpose is to \textit{\quoted{contain costs}} (R19) or to set the output to the \textit{\quoted{bare minimum to have a fixed response}} (R4). As another researcher summarized, this strategy \textit{\quoted{has helped to limit the usage (and cost) of the models}} (R3).

    \smallskip
    \item \textbf{(M12) Multi-Stage Hybrid Pipelines:} To avoid using expensive LLMs for simple tasks, researchers design multi-stage pipelines that act as a cost-saving filter. A common pattern is to use a fast, low-cost method, such as leveraging vector embeddings for an initial analysis \cite{anandayuvaraj2024fail}. The powerful and expensive LLM is then utilized in a second stage to perform a deeper analysis on the much smaller subset of candidate data that has passed the initial filter, thereby optimizing the use of costly resources.
\end{itemize}
\end{description}

\begin{description}[leftmargin=0.3cm]
\smallskip
\item \faExclamationCircle\  \textbf{T5: Reproducibility \& Output Stability.} The reproducibility of results is threatened by a combination of factors, including the inherent stochasticity of LLMs, the "black-box" nature of proprietary models, and the continuous, often undocumented, evolution of the models themselves. The threat of \textbf{temporal drift}, where proprietary models can yield different results over time despite being called by the same name, makes long-term replication particularly difficult, a threat explicitly noted in the literature \cite{wu2024semantic}. These issues are frequently reported across the reviewed studies \cite{de2024developers, colavito2025benchmarking, zhang2024leveraging, anandayuvaraj2024fail, sheikhaei2024empirical, chatterjee2024shedding}.

While this is a well-documented concern, our survey suggests it is perceived as a less frequent operational hurdle compared to other threats. Only three out of 22 respondents (13.6\%) encounter reproducibility issues \squoted{Often} or \squoted{Always}. The vast majority experience it less frequently, with twelve reporting it \squoted{Sometimes} and seven \squoted{Rarely}. Notably, no respondent claimed to have \squoted{Never} faced this threat. However, the difficulty in achieving stable results remains a practical concern. To combat output instability, a mitigation strategy mentioned by numerous respondents is to set the model's \textbf{temperature} parameter to zero. As one researcher stated, their goal is to set parameters \textit{\quoted{such that the responses are deterministic}} (R4). This suggests that practitioners are actively configuring models to mitigate the risks associated with stochasticity, a key component of the broader reproducibility problem.

\smallskip
\noindent\textbf{Mitigation Strategies.} To address the threats of non-determinism, improve the stability of their findings, and improve reproducibility, researchers could employ three strategies:
\begin{itemize}
    \smallskip
    \item \textbf{(M13) Deterministic Parameter Configuration:} The most widely adopted mitigation, reported in both the literature \cite{colavito2025benchmarking, anandayuvaraj2024fail} and our survey, is to configure the LLM for deterministic output. This is almost universally achieved by setting the \textbf{\squoted{temperature}} parameter to \squoted{0}. Numerous survey respondents (R1, R7, R12, R17, R19) confirmed this practice, with one specifying they use a lower temperature \textit{\quoted{when I need more deterministic and consistent outputs}}(R17). This strategy is also highlighted in the approach \textbf{A.9}.

    \smallskip
    \item \textbf{(M14) Stabilizing Stochastics through Aggregation:} When non-deterministic outputs are used (i.e., \squoted{temperature > 0}), researchers ensure the stability of the final reported result by aggregating multiple runs. Instead of reporting the outcome of a single, random run, studies perform the same query multiple times and then aggregate the results, either by taking a \textbf{majority vote} on the outcomes~\cite{zhang2024leveraging} or by \textbf{averaging performance scores} across the runs~\cite{wu2024semantic, colavito2025benchmarking}. This approach reduces variance and ensures the overall experimental result is stable and reproducible. The practice is confirmed by a survey respondent who, to handle uncertainty, will \textit{\quoted{run experiments multiple time and use a vote criterion to determine the accuracy}} (R18). This approach is also described in \textbf{A10}.

    \smallskip
    \item \textbf{(M15) Transparent Reporting and Version Pinning:} A crucial step for ensuring reproducibility and mitigating model drift is the meticulous documentation of the experimental setup. This includes transparently reporting all key parameters (temperature, max tokens, etc.) and, critically, pinning the specific model version used (e.g., \squoted{gpt-3.5-turbo-0125} instead of the generic \squoted{gpt-3.5}) \cite{anandayuvaraj2024fail, colavito2025benchmarking}. This practice acknowledges that models evolve and helps other researchers replicate the study conditions as closely as possible. Furthermore, it is worth noting that replication may be easier with older versions of open-source LLMs than with commercial LLMs, as they may no longer be available. This solution is described in \textbf{A.9}.
    
\end{itemize}
\end{description}

\begin{description}[leftmargin=0.3cm]
\smallskip
\item \faExclamationCircle\ \textbf{T6: Output Formatting \& Parsing.}
Ensuring that LLMs produce output in a consistent, machine-parsable format has been identified in the literature as a recurring threat, often requiring manual correction or the development of post-processing scripts \cite{silva2024detecting}. This issue is a practical annoyance confirmed by our survey respondents' descriptions of their automation workflows. Researchers described developing custom scripts to handle inconsistent outputs. For example, one practitioner's pipeline involves a script for \textit{\quoted{checking for structural consistency, format issues, and unexpected characters}} (R6). Another detailed a common fallback strategy for parsing: if the output has the requested structure, it is parsed automatically, \textit{\quoted{otherwise I use regular expressions to find an answer. If it does not work, the output is considered null}} (R15). These accounts highlight the hidden post-processing effort required in LLM-based mining.

\smallskip
\noindent\textbf{Mitigation Strategies.} To mitigate this threat, researchers could employ three strategies:

\begin{itemize}
    \smallskip
    \item \textbf{(M16) Prompt-Based Formatting Instructions:} This consists of explicitly defining the desired output structure within the prompt itself. Many studies instruct the LLM to return its response in a structured, machine-readable format like JSON \cite{jiang2024peatmoss, wu2024semantic}. To further improve compliance, prompts are often augmented with clear reporting instructions or even few-shot examples that demonstrate the exact output format required \cite{de2024developers}. This mitigation aligns with methodological practices \textbf{(A5)}, which emphasize structured prompting and format specification.

    \smallskip
    \item \textbf{(M17) Robust Post-Processing \& Fallback Parsing:} Recognizing that LLMs do not always produce outputs that strictly follow the requested format, researchers develop resilient post-processing pipelines. These pipelines often implement a multi-level parsing strategy, a workflow one respondent detailed as parsing \textit{\quoted{the output automatically (if it has the requested structure); otherwise i use regular expressions to find an answer}} (R15). Other researchers confirm using \textit{\quoted{regular expression for output data extraction}} (R11) as a fallback mechanism~\cite{abedu2025repochat}. Further techniques include flexible text processing to normalize minor variations (e.g., treating \squoted{Confused} and \squoted{Confusion} as equivalent)~\cite{imran2024uncovering}. This mitigation reflects the approach \textbf{(A4)}.

    \smallskip
    \item \textbf{(M18) Manual Correction or Exclusion:} When automated parsing proves insufficient, researchers adopt one of two strategies. The first is manual correction, where malformed outputs are inspected and adjusted by hand to ensure inclusion in the dataset \cite{silva2024detecting}. The second, more scalable alternative is to exclude non-compliant outputs from the analysis entirely, with the discard rate reported transparently as a limitation of the method \cite{colavito2025benchmarking}.
\end{itemize}
\end{description}

\begin{description}[leftmargin=0.3cm]
\smallskip
\item \faExclamationCircle\  \textbf{T7: Dataset Noise, Imbalance, \& Pre-training Leakage.}
The quality of the data used to prompt LLMs and the data on which they were pre-trained are critical, as highlighted in the literature through threats like annotation noise \cite{sheikhaei2024empirical, shafikuzzaman2024empirical}, dataset imbalance \cite{zhang2025revisiting, kim2025exploring}, and data leakage from pre-training corpora \cite{colavito2025benchmarking, jiang2024peatmoss}. These threats were also raised by our survey respondents. In their open-ended responses, practitioners explicitly identified \textit{\quoted{Bias due to data leakage}} (R19) and the general need for \textit{\quoted{Data quality control}} (R22) as key limitations. This highlights a shared awareness of the \squoted{garbage in, garbage out} principle, where the quality of both the input prompts and the LLM's training data fundamentally constrains the validity of the results.

\smallskip
\noindent\textbf{Mitigation Strategies.} To address data-related threats, researchers focus on rigorous curation of ground-truth datasets, prompt-based techniques to steer model behavior, and and could follow three mitigations:
\begin{itemize}
    \smallskip
    \item \textbf{(M19) Input Dataset Preparation \& Cleaning:} To ensure that the LLM operates on accurate and unbiased data, researchers carefully prepare and clean the input datasets before running the model \cite{de2024developers}. This typically involves using multiple human annotators and conducting reconciliation sessions to resolve disagreements and build a consistent gold standard \cite{imran2024uncovering}. In addition, several studies perform manual error analysis on LLM outputs to identify and retrospectively correct mislabeled examples in the original dataset \cite{sheikhaei2024empirical, zhang2025revisiting}. 

    \smallskip
    \item \textbf{(M20) Bias Mitigation through Prompting \& Analysis:} While controlling the bias inherent in pre-trained models is difficult, one novel mitigation strategy from the literature attempts to steer the model at inference time. This involves embedding safeguards directly into the prompt, such as assigning the LLM an \quoted{unbiased} persona, to guide it away from generating biased or socially harmful content \cite{kim2025exploring}. Complementing this, one survey respondent suggested a more analytical approach: to \textit{\quoted{apply interpretability methods to explain the LLM behavior}} (R22), which can help uncover and understand the sources of bias in a model's decisions. This approach is also highlighted in (\textbf{A.6})

    \smallskip
    \item \textbf{(M21) Acknowledgment of Data Leakage:} Data leakage from the LLM’s pre-training corpus—particularly in proprietary closed weight models, is often an unresolvable threat. In these cases, researchers cannot verify whether the model has seen the target data during training, which undermines construct validity. As a result, the most appropriate mitigation is to explicitly acknowledge this risk in the study’s design and limitations. This includes transparently stating the potential for data leakage and advising readers to interpret findings with appropriate caution~\cite{colavito2025benchmarking, jiang2024peatmoss}. Doing so reinforces the study's integrity and supports informed interpretation, even when the threat cannot be fully eliminated.

\end{itemize}
\end{description}

\begin{description}[leftmargin=0.3cm]

\smallskip
\item \faExclamationCircle\  \textbf{T8: Validation Complexity.} The literature notes that validating LLM outputs is complex and requires significant manual effort to establish a ground truth, especially for nuanced tasks \cite{nguyen2025majority, kim2025exploring}. This point is one of the most strongly supported findings from our survey, highlighting validation as a critical bottleneck in practice. A significant majority of our respondents (13 out of 22, or 59.1\%) report that validation is \squoted{Often} or \squoted{Always} a complex and time-consuming process. An additional five respondents encounter this \squoted{Sometimes}, and four find it to be a \squoted{Rare} issue. Notably, no researcher claimed to have \squoted{Never} faced validation complexity. This complexity was frequently mentioned as the most significant threat. One researcher attributed it to \textit{\quoted{the need for a statistically significant oracle, which required manual curation of the dataset}} (R6). Another one added a crucial perspective on the difficulty of automating this step, stating: \textit{\quoted{we still could not find reliable sources to demonstrate that validating LLM results with LLMs is scientifically sound}} (R21). These insights underline the current necessity of human-in-the-loop validation in empirical SE studies involving LLMs.

\smallskip
\noindent\textbf{Mitigation Strategies.} Researchers mitigate validation complexity with two primary human-centric strategies that trade rigor for scale.
\begin{itemize}
    \smallskip
    \item \textbf{(M22) Oracle Construction for Output Evaluation:} To rigorously validate LLM predictions, researchers construct high-reliability oracles using structured, multi-annotator workflows. This \quoted{gold standard} approach typically involves: (i) having at least two researchers independently label the same data; (ii) a reconciliation session to resolve disagreements; and (iii) reporting inter-rater reliability metrics (e.g., Cohen's Kappa) to demonstrate the ground truth's consistency \cite{imran2024uncovering, zhang2024leveraging, anandayuvaraj2024fail}. However, survey respondents say that \textit{\quoted{complex or time-consuming validation is due to the need for a statistically significant oracle, which requires manual curation of the dataset.}} (R6). Even if we have described these approaches (\textbf{A7, A11}), future research could develop annotated datasets to improve the use of LLM4MSR.

    \smallskip
    \item \textbf{(M23) Scalable Validation via Statistical Sampling:} To address the validation burden in large-scale studies, researchers adopt a strategy based on statistical sampling. Instead of validating every LLM output, a randomly selected, statistically representative subset is manually reviewed to estimate overall model accuracy. This trade-off enables performance assessment while minimizing cost and effort. As noted by our survey respondents, researchers often \textit{\quoted{use sampled data and statistically validate the outputs}} (R10), or more succinctly, \textit{\quoted{only sample statistical sample for evaluation}} (R16) when full validation is not feasible.

\end{itemize}
\end{description}

\begin{description}[leftmargin=0.3cm]
\item \faExclamationCircle\  \textbf{T9: Lack of Standardized Tooling \& Infrastructure.} In addition to methodological threats, researchers frequently encounter a lack of standardized tooling tailored for LLM-based MSR studies. The current landscape consists primarily of low-level APIs and general-purpose libraries, requiring significant engineering effort to operationalize studies. This gap emerged clearly in our survey responses, where participants repeatedly cited the need for \quoted{tools} (R3), \textit{ \quoted{workflow tools to automate the process}} (R7), and \textit{ \quoted{better frameworks/API}} (R4). As a result, researchers are often forced to build bespoke solutions from scratch—an effort described as \textit{\quoted{tremendously time consuming}} (R4), particularly when dealing with issues such as GPU memory and library incompatibilities. These infrastructure issues divert time and resources away from core research objectives, introducing significant delays in the research pipeline.

\smallskip
\noindent\textbf{Mitigation Strategies.} In response to the absence of mature automation tools, researchers adopt engineering-intensive mitigation strategies. We identified three mitigation strategies that directly correspond to the approach (\textbf{A4}), where researchers establish the necessary infrastructure to execute their work.
\begin{itemize}
    \smallskip
    \item \textbf{(M24) Developing Custom Automated Pipelines:} The most direct mitigation is to build the missing tools themselves \cite{de2024developers}. Researchers report creating custom \textit{\quoted{Python scripts, and... an automated pipeline that runs all the phases}} (R2) to manage everything from data extraction and prompt execution to output validation. These pipelines often include custom logic for parsing inconsistent outputs, such as using regular expressions as a fallback when structured formatting fails (R15), thereby codifying the entire experimental workflow for transparency and reproducibility.

    \smallskip
    \item \textbf{(M25) Integrating Libraries \& APIs:} Instead of building silos, researchers act as integrators, stitching together traditional MSR libraries with LLM APIs. For instance, they combine data extraction using established tools like \textit{\quoted{PyDriller or simple GraphQL queries}} (R3) with calls to various LLM provider APIs, creating a functional but often brittle end-to-end workflow.
    
    \smallskip
    \item \textbf{(M26) Adopting Local Inference Engines:} To overcome the financial and accessibility barriers of proprietary API-based infrastructure, a growing mitigation is to set up local inference environments. Researchers use frameworks like \textit{\quoted{llama.cpp}} (R9) or \textit{\quoted{ollama}} (R15) to run open-source models on their own hardware. While this reduces API costs and increases control, it trades one threat for another, requiring significant effort in system configuration and hardware management.
\end{itemize}
\end{description}

\definitionbox{RQ$_2$ — Large Language Model Threats}{As a result of \textbf{RQ$_2$}, we identified nine threats that researchers encounter when using LLM-based MSR. These threats include hallucinations, prompt sensitivity, validation complexity, limited reproducibility, and high computational cost. To address these threats, we analyzed the literature and survey responses to derive 26 mitigation strategies. These include techniques such as hybrid validation workflows, structured prompt engineering, cost-performance-aware model selection, and intelligent context management. By mapping each threat to a corresponding strategy, our findings provide a structured basis for guiding future researchers in selecting informed methodologies.}

%% file: Section/Discussion.tex
\section{Discussions and Implications}
\label{sec:discussion}
Our study offers several implications for different stakeholders involved in integrating LLMs into MSR studies. In the following, we outline the key implications for the SE research community, with the goal of supporting researchers and practitioners in designing more rigorous studies by addressing possible threats to validity by design. By synthesizing current practices and threats, our results provide actionable insights that can guide future empirical work involving LLMs and contribute to advancing the methodological robustness and reproducibility of MSR studies.

\subsection{Understanding Emerging Approaches in LLM-based MSR}
The findings from \textbf{RQ$_1$} shed light on how SE researchers are currently operationalizing LLMs for MSR, offering a resource for those aiming to assess the gap between high-level methodological guidelines and the actual empirical practices adopted in recent LLM-based MSR studies.
These results enrich the existing literature in a way that has not yet been fully captured by prior literature. Hou et al.\cite{hou2024large} provide an extensive mapping of LLM applications in SE and identify mining insights from platforms such as GitHub and StackOverflow as an application area. However, their work does not examine how these mining tasks are practically implemented in research workflows. Similarly, Fan et al. \cite{fan2023large} and Zheng et al.\cite{zheng2025towards} focus on what LLMs can do, offering taxonomies of tasks and capabilities, but they do not analyze how these models are used in practice. Our findings fill this gap by uncovering the decisions and adjustments that researchers make when integrating LLMs into software repository analysis, which are implicit or omitted in publications. The practicalities of prompt formulation, model configuration, and validation are foundational to the success of such studies, yet remain under-discussed.

This absence of methodological transparency has been explicitly noted in recent work. Sallou et al. \cite{sallou2024breaking} warn of the threats posed by the uncritical use of LLMs in SE, highlighting the need for greater reflection on assumptions and potential biases. Wagner et al. \cite{wagner2024towards} respond by calling for evaluation guidelines to ensure the rigor of LLM-based empirical studies. Trinkenreich et al. \cite{trinkenreich2025get} similarly emphasize the growing role of LLMs in SE research, urging the community to take a more systematic approach to experimentation. Lastly, De Martino et al. \cite{de2025framework} create a preliminary framework for utilizing LLMs in MSR studies, focusing on creating, refining, and validating prompts that enhance LLM output, particularly in the context of data collection in empirical studies. While these papers articulate broader concerns, our study provides concrete empirical evidence of the kinds of methodological practices that are emerging, as well as the lack of standardized ways to report or reuse them.

The contrast between the operational depth observed in our analysis and the limited methodological framing in prior work suggests that the community is undergoing a quiet methodological shift, one that is not yet reflected in common standards or publication norms. To support the future of LLM-based MSR, and general SE studies, it will be necessary to move beyond task taxonomies and performance benchmarks and begin investing in shared methodological assets such as prompt libraries, configuration templates, validation heuristics, and reporting guidelines that can enable reproducible, transparent, and scalable research described in our approach \textbf{A15}.

\implication{1}{The findings from \textbf{RQ$_1$} reveal a growing methodological maturity in how LLMs are employed for MSR, but this evolution remains under-documented and fragmented. To support robust and reproducible research, we call for community efforts to consolidate prompting practices, configuration guidelines, and validation heuristics into standardized, reusable methodological resources. Lastly, these approaches could be generalized to other areas of software engineering research.}

\subsection{Revisiting Technical and Methodological Threats To Validity in MSR}

The findings from \textbf{RQ$_2$} provide a deeper understanding of the obstacles researchers face when integrating LLMs into MSR. These threats span from the technical to the methodological and are not always acknowledged explicitly in the literature. In our analysis, researchers reported difficulties in selecting suitable LLMs for their tasks, tuning them effectively under resource constraints, and validating their outputs in the absence of ground truth data. These issues were particularly pronounced when dealing with under-documented APIs, hallucinated outputs, and inconsistent results between runs. Moreover, researchers often had to make trade-offs between automation and manual oversight, balancing efficiency against control and interpretability. Many of these decisions were shaped by the limitations of current tooling and the lack of standardized evaluation practices.
To enrich the practical relevance of these findings, one of the key contributions of our work is that we explicitly annotated each threat with its associated mitigation strategies. No study has classified the threats posed by LLMs to MSR studies and defined mitigation strategies to support research. These include, for example, the risk of unstable outputs due to non-deterministic decoding settings, which researchers addressed by enforcing fixed seeds and low-temperature configurations. Similarly, the threat of prompt misalignment, where minor changes in phrasing affect outcomes, was mitigated through iterative pilot testing. By capturing these concrete mitigation strategies, our analysis provides actionable guidance for future researchers. It offers a framework for reasoning about and mitigating the methodological fragility of LLM-based usage in MSR studies. Furthermore, these threats are likely to be relevant to areas other than MSR. Although this empirical study focuses on MSR, threats have been identified that relate to the general use of LLMs outside MSR (e.g., hallucinations), and these threats, mitigations, and approaches have direct implications for the broader use of LLMs.

Our findings align with recent discussions on how the software engineering community addresses threats to validity. Lago et al. \cite{lago2024threats} emphasize that these threats are often underreported or treated as formalities, urging a move beyond standardized checklists to better understand their emergence and management in research. Our investigation focuses on how technical and methodological threats occur in studies involving LLMs in MSR. While some identified issues, like construct misalignment, fit traditional validity categories, others, such as unstable decoding behaviors and prompt sensitivity, arise specifically from LLMs. These insights underscore the need for both taxonomical clarity and practical understanding of how researchers navigate uncertainty and adapt their methods. Our work complements Lago et al. \cite{lago2024threats} agenda by offering a domain-specific perspective on threats to validity in LLM-based research.

Additionally, these threats reinforce and refine observations previously made in the literature. Sallou et al. \cite{sallou2024breaking} have warned that the adoption of LLMs in SE is often accompanied by a lack of critical reflection, with risks stemming from opacity, reproducibility issues, and fragile assumptions. Wagner et al. \cite{wagner2024towards} propose evaluation guidelines to improve the quality of empirical studies involving LLMs. Our findings empirically validate these concerns: even in well-structured studies, we found limited reporting on validation protocols, replication attempts, or the rationale behind LLM selection. Hou et al. \cite{hou2024large}, in their SLR, identified multiple research opportunities related to LLM integration, scalability, and data management, but their analysis stops short of detailing the technical and methodological frictions that occur in practice. 

These insights suggest that the integration of LLMs into MSR research is not only a matter of LLM capability, but also of process maturity. Methodological fragmentation, lack of reusable assets, and poor tooling support create barriers to scalability, reproducibility, and generalization. This is particularly concerning in a field like MSR, where the quality of insights depends on the consistency and traceability of the mining process. The threats identified in \textbf{RQ$_2$} reveal that SE researchers, in the pioneering LLM-based MSR studies (LLMs gained popularity in November 2022), have been left to make ad hoc decisions, constrained by limited computational resources, unstable APIs, and inadequate documentation.

To advance the field, we propose that issues such as \quoted{Contextual Understanding \& Fomain Dpecificity} (T3), \quoted{Validation Complexity} (T8), and \quoted{Reproducibility \& Output Stability}(T5) should be viewed not only as technical challenges but also as methodological research problems. Addressing these issues requires coordinated efforts from the community. This could involve creating open-source datasets with gold-standard annotations to tackle dataset noise (T7) and validation complexity (T8), developing streamlined validation protocols as a direct response to validation uncertainty, designing model comparison toolkits specifically for SE tasks to address output inaccuracy (T1) and output instability (T5), and establishing evaluation frameworks that extend beyond traditional metrics like accuracy or F1 score. 

\implication{2}{The findings from \textbf{RQ$_2$} show that the main barriers to progress in LLM-based MSR are not purely technical, but methodological. By mapping threats and mitigation strategies for each identified threat, our study offers actionable knowledge that can inform the design of more stable and reproducible LLM-based pipelines. The MSR and general SE fields now need shared protocols, lightweight validation strategies, and infrastructure for prompt and model management to support robust, scalable, and replicable research practices.}

\subsection{Implications for the Mining Software Repositories Community}
The rise of LLMs for software repository analysis, and especially for data collection and analysis, represents both an opportunity and a challenge for the MSR community. LLMs promise to enhance traditional mining activities, such as classification and entity extraction, beyond the limits previously achievable with rule-based or shallow learning techniques. However, there is still a lack of a common methodological foundation, resulting in fragmented practices and reproducibility issues that hinder the proper conduct of such studies.

From a community development perspective, this situation calls for a change in the way LLM-based studies are designed, reported, and reviewed. Researchers are no longer mere consumers of pre-trained models, but are actively shaping the behavior of LLMs through prompts, tuning, and post-processing. This change implies that the MSR community, to continue being a reference about open science and reproducibility, must now include in its pipeline-oriented discipline new artifacts as first-order research contributions: prompt design, validation strategies, and reproducibility protocols. In fact, our study shows that many SE researchers are already engaged in this type of methodological work.

Looking ahead, we believe that in the MSR and SE communities, we should actively invest in creating methodological guidelines to support the design and evaluation of LLM-based studies. These guidelines should address essential aspects such as prompt formulation, configuration control, reporting practices, and validation protocols. Similar to past community efforts in dataset curation and shared activities, these resources could foster more robust and transparent research, enabling others to replicate results, compare techniques, and build on previous work in a cumulative manner.

Furthermore, as LLMs become increasingly central to empirical SE, the community will face fundamental challenges related to reproducibility. Unlike traditional mining pipelines, LLM-based workflows often involve stochastic outputs and dependence on rapidly evolving model versions, all of which pose a threat to long-term replicability. Ensuring the reproducibility of research studies will require not only technical solutions (e.g., versioned models, deterministic decoding), but also the extension of more detailed documentation and open sharing of prompts, code, and inference artifacts as described in (\textbf{A15}).

More broadly, the approaches in \textbf{RQ$_1$} and the challenges in \textbf{RQ$_2$} extend not only to the MSR community but also to the SE community. As LLMs are increasingly used to generate explanations, summaries, and retrieve artifacts, SE researchers need to critically examine what constitutes evidence, how it should be evaluated, and what ethical responsibilities arise from automating human-level reasoning tasks.

\implication{3}{The integration of LLMs into MSR research opens up new opportunities to go beyond extracting data specific to certain tasks and move toward comprehensive experimental design. To respond to this change, the community must develop a shared infrastructure, define methodological standards, and explicitly address challenges related to reproducibility and validation. This will ensure that LLM-based MSR studies are not only innovative, but also rigorous, transparent, and sustainable over time.}

\subsection{PRIMES 2.0: An Empirical-Based Methodological Framework for LLM-Based Mining of Software Repositories}

Building on the results of \textbf{RQ$_1$} and \textbf{RQ$_2$}, we propose an evolved version of our initial framework, PRIMES \cite{de2025framework}, aimed at supporting the design and execution of LLM-based MSR studies. The original PRIMES framework had some limitations, as it was based on only two previous studies\cite{de2024developers,castano2024machine} and did not fully address the complexities and threats encountered in the full research pipeline. While the initial version of PRIMES offered a conceptual framework for configuring prompts and models, our study highlighted the need for enhancements in two key areas: methodological improvement and the integration of strategies for validating and mitigating potential threats.

\begin{figure}[h]
    \centering
    \includegraphics[width=1\linewidth]{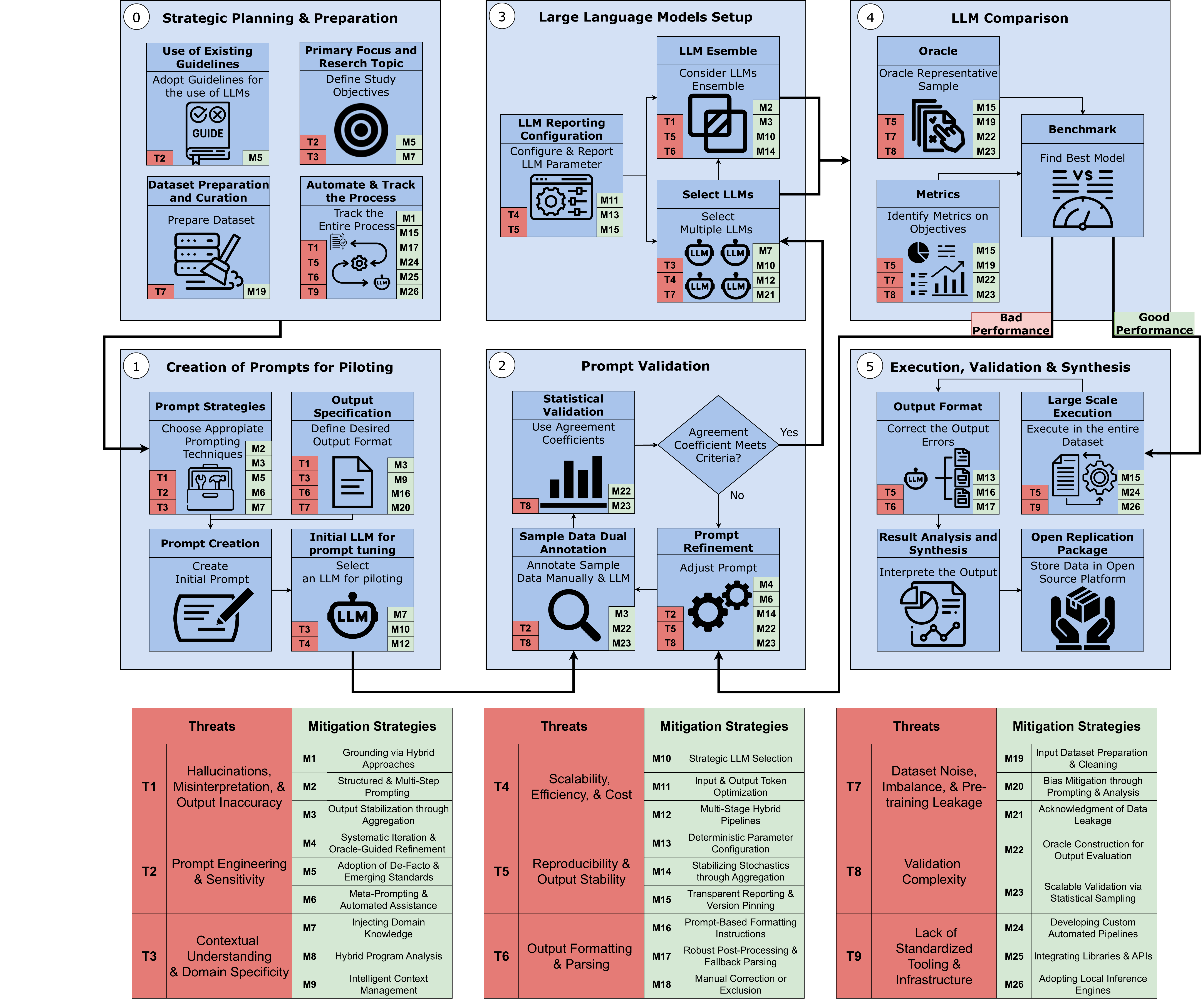}
    \caption{PRIMES 2.0: An Empirical-Based Methodological Framework for LLM-Based Mining of Software Repositories.}
    \label{fig:primes}
\end{figure}

The new PRIMES 2.0, shown in Figure \ref{fig:primes}, addresses the limitations identified in earlier frameworks by incorporating findings from our mixed-method study. PRIMES 2.0 is organized into six stages, comprising 23 methodological substeps, each mapped to specific threats (T1–T9) and corresponding mitigation strategies (M1–M26), providing prescriptive and adaptive support throughout the lifecycle of LLM-based MSR studies.

PRIMES 2.0 functions as both a design tool and a diagnostic instrument. It empowers researchers to make informed decisions at each stage of the process, from selecting representative repositories and designing reproducible prompts to conducting large-scale evaluations and interpreting results. Unlike traditional linear methodologies, PRIMES 2.0 supports iterative refinement loops, allowing prompts, model configurations, and validation strategies to evolve based on empirical feedback and pilot results. This adaptability is particularly important in MSR contexts, where prompt behaviors can vary significantly depending on the repository context, historical depth, and task-specific semantics. PRIMES 2.0 builds upon and extends previous guidelines tailored for LLM-based software engineering research by offering a structured, staged framework that spans from \quoted{strategic planning \& preparation} to \quoted{execution, validation \& synthesis}. It is a six-stage process that formalizes best practices for both the design and execution of empirical studies involving LLMs.

Wagner et al.~\cite{wagner2024towards} offer a rich set of practical recommendations aimed at improving the transparency, reproducibility, and interpretability of LLM-based SE studies. These include documenting model versioning and decoding parameters, prompt structure and rationale, evaluation criteria, and annotation agreement practices, all of which are also reflected in the studies we analyzed. Our findings corroborate these guidelines and reinforce the emerging consensus on best practices for LLM-based research. Our contribution builds on this work by explicitly mapping such practices to recurring methodological threats observed in the field, and by embedding them into a structured, multi-step protocol. In this sense, PRIMES 2.0 complements previous efforts by operationalizing and organizing these recommendations into concrete stages, mitigation strategies, and validation techniques (e.g., dual annotation, LLM-as-judge loops, inter-model comparisons), thus enabling more fine-grained methodological traceability.

Additionally, PRIMES 2.0 incorporates sustainability aspects from both environmental and methodological perspectives. Expanding on the reflections introduced in the original framework \cite{de2025framework}, PRIMES 2.0 incorporates sustainability into several stages of the empirical workflow. Specifically, we address the high computational cost and resource consumption associated with large-scale LLM usage by introducing scalable validation techniques and model reuse strategies. For example, rather than evaluating every prompt-model combination exhaustively, PRIMES 2.0 promotes validation via representative sampling and focuses only on configurations that show initial evidence of reliability. The framework encourages reuse of models through local inference setups with open-weight LLMs and supports parameter stabilization to avoid redundant executions and uncontrolled variability across runs.
Moreover, PRIMES 2.0 reinforces reproducibility and traceability by requiring researchers to document prompt templates, versioned model configurations, and output logs. These artifacts are designed to be reusable across studies, reducing unnecessary repetition and promoting knowledge sharing. In line with prior concerns about private models and pretraining leakage~\cite{sallou2024breaking}, the framework prioritizes transparent model selection and encourages the use of open-source resources. By embedding these practices directly into the methodological design, PRIMES 2.0 responds to the need for energy-aware, ethically grounded, and methodologically sound LLM-based research in software engineering.

In addition, the work by Sallou et al.~\cite{sallou2024breaking} raises awareness of construct, internal, and external validity threats, especially for LLMs trained on public datasets. Their guidelines focus on data leakage detection, reproducibility via prompt logs and date stamps, and the use of metamorphic testing. PRIMES 2.0 concretely operationalizes these principles by (i) introducing a dual-annotation process for prompt validation, (ii) recommending metamorphic input variants during benchmarking, and (iii) requiring execution metadata and output archiving across all stages. It also reinforces traceability through integrated LLM versioning and automates and tracks the entire process, aligning with the concerns highlighted by Sallou et al. ~\cite{sallou2024breaking}.

Although PRIMES 2.0 was developed from empirical studies within the MSR community, its structure and practices are not unique to this domain. While originally designed to support mining-specific tasks, the framework formalizes methodological decisions such as prompt refinement, model comparison, and output validation that are broadly relevant across empirical SE.
Several components of PRIMES 2.0 are applicable to other SE contexts where LLMs are utilized for analysis, synthesis, or labeling, including test generation, code summarization, and requirements classification. In these domains, researchers face similar methodological threats, such as prompt sensitivity, output reproducibility, and lack of validation infrastructure, which PRIMES 2.0 directly addresses through its threat-driven design. We envision PRIMES 2.0 as a foundation for future guidelines, protocols, and toolkits that support not only MSR studies but also the broader empirical SE community. Its modular and extensible structure encourages researchers to report both outcomes and the processes that led to them. We invite the community to explore, apply, and extend PRIMES 2.0 to assess its applicability across diverse empirical settings.

\implication{4}{PRIMES 2.0 operationalizes the insights from our empirical study into a flexible and threat-aware framework that supports rigorous design, refinement, and validation of LLM-based MSR studies. It provides a methodological foundation consisting of mitigation actions for building reproducible, transparent, and adaptive empirical pipelines.}

%% file: Section/Threats.tex
\section{Threats to Validity}
\label{sec:threats}
This study, which combines a rapid literature review with a questionnaire survey study, is subject to several validity threats. In this section, we discuss potential limitations and describe the strategies employed to mitigate them \cite{wohlin2012experimentation}.

\subsection{Construct Validity}

Construct validity refers to the extent to which our methods and instruments accurately capture the intended constructs \cite{wohlin2012experimentation}. In our case, a primary concern was whether the survey questions and review criteria effectively captured researchers’ practices and threats in applying LLMs to repository mining. 
To address this, we followed questionnaire survey design guidelines by Dillman et al. \cite{dillman2014internet} and Kitchenham et al. \cite{kitchenham2008_PersonalOpinionSurveys} to ensure the clarity and alignment of the questionnaire survey instrument with our research questions. Additionally, we conducted a pilot test with four SE researchers experienced with at least one year of experience with the use of LLM in software repository mining. Their feedback helped us refine question phrasing, remove redundancies, and estimate completion time. We also embedded an attention-check question to filter careless responses, helping to preserve the reliability of the collected data.

In the rapid review, construct validity risks arise from the interpretation of what constitutes LLM use in software repository mining. To address this, we operationalized inclusion and exclusion criteria inspired by Hou et al. \cite{hou2024large} and focused exclusively on peer-reviewed studies. These criteria were applied systematically to ensure consistency with the goals of the study.

\subsection{Internal Validity}

Internal validity concerns the credibility of the observed patterns and the minimization of confounding factors \cite{wohlin2012experimentation}. One potential threat in our study stems from the subjective interpretation of open-ended questionnaire survey responses. To mitigate this, we employed qualitative content analysis \cite{krippendorff2018content}. The first and second authors independently reviewed the qualitative responses and iteratively developed thematic categories through collaborative discussion. This minimized the risk of misinterpretation due to individual bias. In the rapid review, internal validity risks relate to the selection of primary studies. To mitigate subjective bias during the inclusion process, the first and second authors independently reviewed all papers selected from the seed set. Disagreements were resolved through discussion. To assess reviewer agreement, we computed Cohen’s Kappa \cite{cohen1960coefficient} in two consecutive rounds. After a consensus session to clarify ambiguities and achieve a high Cohen’s Kappa, we proceed to analyze the articles. This procedure helped ensure a consistent and reliable application of inclusion and exclusion criteria.

\subsection{External Validity}

External validity addresses the extent to which our findings can be generalized beyond the context of our study \cite{wohlin2012experimentation}. For the questionnaire survey, we targeted researchers with demonstrable experience using LLMs for software repository mining. Participants were recruited through a combination of purposive and convenience sampling strategies, including academic networks, LinkedIn posts, and direct invitations to authors of papers included in the review. Although this limits statistical generalizability, it enabled us to reach an expert population and collect meaningful, experience-based insights \cite{baltes2022_convenience_sampling_SE}.

The rapid review was constrained to high-quality software engineering venues, in line with standard practice for ensuring relevance and methodological rigor \cite{wohlin2020guidelines}. However, this decision may have excluded relevant articles from other conferences or niche venues. To address this, we conducted a two-level snowballing phase, which allowed us to capture relevant studies that may not have been included in our initial seed set.

\subsection{Conclusion Validity}

Conclusion validity refers to the soundness of the inferences we drew from the data \cite{wohlin2012experimentation}. In the questionnaire survey, we combined quantitative analysis of structured responses with qualitative content analysis of open-ended answers. The consistency between both types of responses and the triangulation with findings from the rapid review increased our confidence in the validity of the conclusions. To further ensure reliability, we filtered out responses that failed the attention-check question or were inconsistent, and manually reviewed each retained submission. 

In the review, our inclusion and exclusion criteria, quality screening, and venue selection increased the methodological transparency and replicability of our process. All decisions and extracted data were documented and made publicly available in our online appendix \cite{appendix}, enabling replication and follow-up studies.

By synthesizing the results of both studies independently and comparing their results, we achieved methodological triangulation \cite{runeson2009guidelines}, which increased the robustness of our interpretations without requiring full data integration, as in mixed-method research.

%% file: Section/Conclusion.tex
\section{Conclusion}
\label{sec:conclusion}
This paper presents a multi-method empirical study on the use of LLMs for MSR. We combined a rapid literature review and a questionnaire survey with researchers to gain insight into both published practices and real-world usage. Our work focuses on two main objectives: identifying the methodological approaches, prompting techniques, and validation strategies currently in use, and identifying the threats researchers face when adopting LLMs in MSR studies and their mitigations. Our work provided the following contributions:

\begin{enumerate}
    \item A rapid review and a questionnaire survey that jointly explore the 14 approaches and actions to mitigate 9 identified threats in applying LLMs to MSR. Together, these studies provide a structured overview of the field by categorizing LLM models, prompting strategies, validation techniques, and evaluation metrics, while also uncovering threats, gaps, and methodological inconsistencies that were not previously captured in the literature.

    \smallskip
    \item PRIMES 2.0, an improved and empirically grounded framework that consolidates emerging practices and mitigation strategies into a structured process, It provides a methodological foundation for the community, supporting the design, execution, and validation of LLM-based MSR studies, and aims to serve as future guidelines for reproducible LLM4SE research. 

    \smallskip
    \item A set of implications and open questions to guide future research, as well as a curated dataset of reviewed papers and anonymized questionnaire survey responses to foster transparency, reproducibility, and reuse.

    \smallskip
    \item A public replication package \cite{appendix} which may be exploited by researchers to either replicate our work or build on top of our findings.
    
\end{enumerate}

The results of this study highlight the fragmented nature of current practices and reinforce the need for methodological guidance in integrating LLMs into MSR workflows. As part of this work, we extended the PRIMES framework by incorporating empirical evidence from both the literature and questionnaire survey, offering a structured basis for designing, executing, and validating LLM-based MSR studies. Future research will focus on validating this extended framework across diverse mining scenarios and research contexts, with the goal of assessing its effectiveness in improving study design, reproducibility, and empirical robustness. We also plan to complement our findings with interviews, replication studies, and more granular analyses of PRIMES 2.0 behavior in other scenarios to uncover task-specific limitations and emerging practices. We encourage other researchers to adopt and contribute to PRIMES 2.0 by sharing prompt templates, validation protocols, and tooling assets.

Finally, considering the increasing importance of generative AI and LLMs in software engineering, future studies may further explore domain-specific issues such as prompt engineering strategies, hallucination mitigation, and sustainability trade-offs in large-scale and real-world settings.